\documentclass[twocolumn,prd,nofootinbib,superscriptaddress]{revtex4-2}
\usepackage{color}
\usepackage{amsmath}
\usepackage{mathrsfs}
\usepackage{graphicx}
\usepackage{booktabs}
\usepackage{bm}
\usepackage{nicematrix}
\usepackage{makecell}
\usepackage{adjustbox} 

\newcommand\mnras{MNRAS}
\newcommand\prx{PRX}
\newcommand\prr{PRR}
\newcommand\apjl{ApJL}
\newcommand\apjs{ApJS}

\newcommand\SkipNeuralNets[1]{}
\newcommand\skipme[1]{}

\newcommand\qmstateproduct[2]{\left\langle#1|#2\right\rangle}

\newcommand\Y[1]{{{}_{#1}Y}}

\newcommand\optional[1]{}
\newcommand\unit[1]{{\rm #1}}

\newcommand\response[1]{{#1}}
\newcommand\mc{{{\cal M}_c}}

\def\RIT{Center for Computational Relativity and Gravitation, Rochester Institute of Technology, Rochester, New York
  14623, USA}

\def\Vand{Department Physics and Astronomy, Vanderbilt University, 2301 Vanderbilt Place, Nashville, TN, 37235, USA}

\def\INFN{INFN Sezione di Torino, Via P. Giuria 1, 10125 Torino, Italy}
\def\Glasgow{SUPA, School of Physics \& Astronomy, University of Glasgow,
  G128QQ}
\def\Birmingham{Institute of Gravitational Wave Astronomy, School of Physics and Astronomy, University of Birmingham, Edgbaston, Birmingham B15 2TT, UK}

\begin{document}

\title{Narrowing RIFT: Focused  simulation-based-inference for interpreting  exceptional GW sources}
\author{K. Wagner}
\affiliation{\RIT}
\author{R. O'Shaughnessy}
\affiliation{\RIT}
\author{A. Yelikar}
\affiliation{\RIT}
\affiliation{\Vand}
\author{N. Manning}
\affiliation{\RIT}
\author{D. Fernando}
\affiliation{\RIT}
\author{J. Lange}
\affiliation{\INFN}
\author{V. Tiwari}
\affiliation{\Birmingham}
\author{A. Fernando}
\affiliation{\RIT}
\author{D. Williams}
\affiliation{\Glasgow}
\begin{abstract}
The Rapid Iterative FiTting (RIFT) parameter  inference algorithm  provides a simulation-based-inference approach to  efficient, highly-parallelized
parameter inference for GW sources.    Previous editions of RIFT have conservatively optimized for robust inference about
poorly constrained observations.  
In this paper, we summarize algorithm enhancements and operating point choices to enable inference for more exceptional
compact binaries.  Using the previously-reported RIFT/asimov interface to efficiently perform analyses on events with reproducible
settings consistent with past work, we demonstrate that the latest version of RIFT can efficiently analyze events with
multiple costly models including the effects of precession or eccentricity.
\end{abstract}
\maketitle


\section{Introduction}
Ground-based  gravitational wave (GW) detectors in the International Gravitational Wave observatory Network (IGWN), including
 Advanced LIGO  \cite{2015CQGra..32g4001L}  and Virgo \cite{gw-detectors-Virgo-original-preferred,2015CQGra..32b4001A},
\response{now joined by KAGRA \cite{2021PTEP.2021eA101A}}  continue to identify
  coalescing compact binaries
\cite{DiscoveryPaper,LIGO-O1-BBH,LIGO-GW170817-bns,LIGO-O3-NSBH,LIGO-O3-O3b-catalog,LIGO-O3-O3a_final-catalog}.
Their  properties can be characterized via Bayesian inference,  comparing data to the expectations given  different potential sources
\cite{DiscoveryPaper,LIGO-O1-BBH,2017PhRvL.118v1101A,LIGO-GW170814,LIGO-GW170608,LIGO-GW170817-bns,LIGO-O2-Catalog,gwastro-PE-AlternativeArchitectures,gwastro-PENR-RIFT,gw-astro-PE-lalinference-v1,ligo-230529}.
The Rapid Iterative FiTting (RIFT) \cite{gwastro-PENR-RIFT}, one of several parameter inference algorithms
\cite{gw-astro-PE-lalinference-v1,gwastro-pe-bilby-2018} used interpret  GW
observations
\cite{LIGO-GW170608,LIGO-GW170817-bns,LIGO-O3-GW190521-discovery,LIGO-O3-GW190521-implications,LIGO-O3-NSBH,LIGO-O2-Catalog,LIGO-O3-O3a-catalog,LIGO-O3-O3b-catalog,LIGO-O3-O3a_final-catalog},
has recently been substantially extended to increase its flexibility and performance \cite{gwastro-RIFT-Update},
culiminating in code releases (0.0.15.7--0.0.15.11) near the start of the O4 observing run.
Unlike the most popular approaches for gravitational wave parameter inference, which rely on Markov chains 
within either Markov Chain Monte Carlo or nested sampling codes (see \cite{RevModPhys.94.025001} for a recent review),
RIFT performs Bayesian inference through a mix of Monte Carlo quadrature; iterative simulation-based inference to
successively approximate pertinent likelihoods; and  considerable  domain-specific expertise (``architectures'') to
target RIFT's iterative refinement.

In this work, we identify limitations and inefficiencies in previous editions of RIFT's recommended operating point and code, albeit only for exceptional
configurations which GW observatories may not soon reveal.  
RIFT's previous highly-conservative approach is particularly ill-suited to sources whose intrinsic and extrinsic
parameters can all be narrowly constrained by inference.   For example, RIFT's domain-specific operating point choices intentionally adopted
simplifications which increase robustness for typical sources:  uniform sampling over several parameters, including distance, polarization, and orbital phase.  Too, motivated by real
 observations like GW190814, a high mass ratio binary consistent with nearly zero spin on the primary, RIFT's
 domain-specific choices emphasized close examination of small spins and particularly small transverse spins.    Though
 well-adapted to interpret previous observations, these choices were suboptimal when presented with highly exceptional
synthetic sources, such as a high-mass ratio binary ($q\simeq 20$) with a strongly-precessing misaligned primary spin,
or certain very strong sources (e.g.,  SNR above 100) where all intrinsic and extrinsic parameters can be well-constrained.

To mitigate these weaknesses, we present architectural and algorithmic advances to RIFT, as well as updates to its default operating point and
infrastructure.  We introduce
new integration techniques; develop new tools to initialize sampling for highly localized sources; and generally add new
domain-specific expertise for high-amplitude, strongly-precessing, or otherwise precisely measurable sources.  We
highlight by example some novel  opportunities that RIFT's heterogeneous architecture fortuitously provides for
self-consistency and validation checks within an analysis, without further postprocessing.
Several of the technical methods developed for this work and first comprehensively described here have already been employed successfully in other work  \cite{gwastro-mergers-rift_asimov_O3-Fernando2024,gwastro-RIFT_LISA-Jan2024,gwastro-mergers-bns-systematics-Yelikar2024,gwastro-ns-eos-pop_systematics-Yelikar2024,gwastro-RIFT-hyperpipe-EOS-Atul,gwastro-ns-eos-Vilkha2024,PhysRevResearch.5.013168,gwastro-mergers-ecc-HectorJake2022,gwastro-pe-eccentric-Wagner2024}.

This paper is organized as follows.
In Section \ref{sec:review}, following \cite{gwastro-RIFT-Update} we review the RIFT algorithm, highlighting
previously-implemented methods described in \cite{gwastro-RIFT-Update} but also describing new interfaces to external
  tools and waveforms not available at the time of that study.
In Section \ref{sec:tests_old_version} we provide anecdotes to illustrate the weaknesses of our previous RIFT
configuration for exceptional sources.  Conversely, by comparison to reanalyses
of GWTC2.1 \cite{LIGO-O3-O3a_final-catalog} and GWTC3 \cite{LIGO-O3-O3b-catalog} events done by
others, we provide anecdotes to demonstrate by example that the previous RIFT configuration produces robust results for
real observations, when given appropriate input settings. 
In Section \ref{sec:previous_work}, we summarize other previously reported extensions to RIFT, highlighting tools used
to validate the improvements described below.
Next, in Section \ref{sec:improve_int} we introduce and validate new Monte Carlo integrator implementations, the primary low-level
domain-agnostic technology underpinning all RIFT calculations.
In Section \ref{sec:technical_internal} we describe other notable changes which impact RIFT's efficiency or robustness.
Section \ref{sec:results} presents new results obtained with the latest RIFT
version and configuration, both on real events using public GW data
\cite{ligo-O1O2-opendata, ligo_o3_opendata} and on synthetic
sets of events (``PP plots'').
Finally, in Section \ref{sec:conclude} we summarize our work, placing it in context with ongoing improvements in GW
parameter inference. 
This paper describes most algorithmic and structural improvements implemented up to and including RIFT version
0.0.17.2.

\section{RIFT review}
\label{sec:review}

A coalescing compact binary can be completely characterized by its intrinsic
and extrinsic parameters.   The intrinsic parameters are necessary to characterize a binary's orbital inspiral and
merger trajectory, up to
spacetime symmetries.  For a quasicircular binary, this trajectory is specified by the binary's  (detector-frame) masses $m_{i,z}$, spins, and any quantities
characterizing matter in the system.  For an eccentric binary, these parameters must be supplemented by the orbital
eccentricity and mean anomaly. 
Conversely, the extrinsic parameters are seven quantities reflecting flat spacetime symmetries: seven numbers needed to characterize its spacetime location and orientation.  
We express masses in solar mass units and
 dimensionless spins in terms of Cartesian components $\chi_{i,x},\chi_{i,y}, \chi_{i,z}$, expressed
relative to a frame with $\hat{\mathbf{z}}=\hat{\mathbf{L}}$ and (for simplicity) at the orbital frequency corresponding to the earliest
time of computational interest (e.g., an orbital frequency of $\simeq 10 \unit{Hz}$).  We will use $\lambda,\theta$ to
refer to intrinsic and extrinsic parameters, respectively.

At a high level of abstraction, RIFT can be understood as a two-stage iterative process to interpret gravitational wave observations $d$ via comparison to
predicted gravitational wave signals $h(\bm{\lambda}, \bm\theta)$.   In one stage,
RIFT computes a marginal likelihood 
\begin{equation}
 {\cal L}\response{({\bm \lambda})}\equiv\int  {\cal L}_{\rm full}(\bm{\lambda} ,\bm\theta )p(\bm\theta )d\bm\theta
\end{equation}
from the likelihood ${\cal L}_{\rm full}(\bm{\lambda} ,\theta ) $ of the gravitational wave signal in the multi-detector network,
accounting for detector response; see  \cite{gwastro-PE-AlternativeArchitectures,gwastro-PENR-RIFT} for a more detailed
specification.  This calculation is performed in parallel on a large number of candidate intrinsic parameters ${\bm
  \lambda}_\alpha$ by a program denoted as ILE (because it integrates the GW likelihood over extrinsic parameters).
In the second stage, RIFT interpolates and integrate.   Specifically, first it interpolates the likelihood information
accumulated from all previous iterations, making an approximation to ${\cal L}(\lambda)$ based on its
accumulated archived knowledge of marginal likelihood evaluations 
$(\lambda_\alpha,{\cal L}_\alpha)$.  Second, using this approximation, it deduces the (detector-frame) posterior distribution
\begin{equation}
\label{eq:post}
p_{\rm post}=\frac{{\cal L}(\bm{\lambda} )p(\bm{\lambda})}{\int d\bm{\lambda} {\cal L}(\bm{\lambda} ) p(\bm{\lambda} )}.
\end{equation}
where prior $p(\bm{\lambda})$ is the prior on intrinsic parameters like mass and spin.

At the end of the iterative calculation, RIFT performs one final pass of ILE over posterior intrinsic draws $\lambda_k$,
fairly drawing some (fixed) number of extrinsic parameters $\theta_{k,p}$ from the  underlying ILE Monte
Carlo integral weighted samples.   These combined samples ($\lambda_k,\theta_{k,p}$) provide a full posterior for all
intrinsic and extrinsic parameters.   Postprocessing each sample provides derived parameters, including the ource-frame masses $m_{i}$.

This section is organized as follows.  In Section \ref{sec:sub:L} we describe how  RIFT evaluates the gravitational wave
likelihood ${\cal  L}_{\rm full}$.  In Section \ref{sec:sub:Lmarg}, we describe how ILE performs Monte Carlo integration
to compute the marginal likelihood ${\cal L}$.   In Section \ref{sec:sub:CIP}, we outline how CIP interpolates the
likelihood, then performs Monte Carlo integration to produce fair draws from the (estimated) posterior distribution.
Then in Section \ref{sec:sub:explore} we explain how RIFT actually uses ILE and CIP iteratively, augmented with
other tools and using binary-inspiral-specific coordinate systems that are well-adapted to GW source parameter
inference.
In Section \ref{sec:sub:pp} we describe  infrastructure used to efficiently validate RIFT code and operating-point
choices with standard PP tests.
In Section \ref{sec:sub:external:Waveforms} we summarize RIFT's updated interface to external waveforms, including key
features needed to ensure compatibility with previous and other work. 
In Section \ref{sec:sub:external:calmarg} we describe selected key external interfaces RIFT uses to provide calibration
marginalization or access new waveforms.
In Section \ref{sec:sub:comments} we briefly comment on RIFT's computational cost and redundancy, highlighting room for improvement on
the one hand and inadvertent but valuable consistency checks on the other.

\subsection{Review of the RIFT likelihood}
\label{sec:sub:L}
As described in previous work \cite{gwastro-PE-AlternativeArchitectures,gwastro-RIFT-Update}, RIFT uses physical
insight to decompose the overall inference calculation  into two stages: inference and marginal likelihoods for a fixed
physical binary, and inference about different physical binaries. 
At a high level, RIFT expresses gravitational wave signals $h(t)$ in terms of physical basis signals $h_{lm}(t)$, associated
with a (spin-weighted) spherical harmonic decomposition of radiation in all possible emission directions.  This
decomposition allows RIFT to compute  cross-correlations between this basis and each detector's data; the likelihood for
arbitrary source orientations,  sky positions, and distances follows by a weighted average of these cross-correlation timeseries.

Using notation established in previous studies
\cite{gwastro-PENR-RIFT,gwastro-PENR-RIFT-GPU,gwastro-PE-AlternativeArchitectures}, the RIFT likelihood is expressed in
terms of a (spin-weighted) spherical harmonic decomposition of the complex gravitational wave strain
\begin{align} \label{eq:strain}
h(t,\vartheta,\phi;\bm{\lambda}) =  h_+(t,\vartheta,\phi;\bm{\lambda}) - 
                                i h_\times (t,\vartheta,\phi;\bm{\lambda}) \, ,
\end{align}
Customarily its real and imaginary parts are denoted its two fundamental polarizations $h_+$ and $h_\times$.
Here, $t$ denotes time, $\vartheta$ and $\phi$ are the polar and azimuthal angles
for the direction of gravitational wave propagation away from the source. 
The complex gravitational-wave strain can be written in terms of
spin-weighted spherical harmonics $\Y{-2}_{\ell m} \left(\vartheta, \phi \right)$ as 
\begin{align} \label{eq:strain_mode}
h(t,\vartheta,\phi;\bm{\lambda}) = 
\sum_{\ell=2}^{\infty} \sum_{m=-\ell}^{\ell} \frac{D_{\rm ref}}{D} h^{\ell m}(t;\bm{\lambda}) \Y{-2}_{\ell m} \left(\vartheta, \phi \right) \, ,
\end{align}
where the sum includes all available harmonic modes $h^{\ell m}(t;\pmb{\bm{\lambda}})$ made available by the model;  where
$D_{\rm ref}$ is a fiducial reference distance; and where $D$, the luminosity distance to the  source, is one of the
extrinsic parameters.  
Each detector has (assumed constant) response functions $F_{+}$ and $F_\times$, such that the time-dependent strain
response has the form  $h_k(t) =F_{+,k} h_+(t_k) +
  F_{\times,k}h_\times(t_k)$ for the detector response $h_k$, 
where $t_k=t_c - \vec{x}_k \cdot \hat{n}$ is the arrival time at the $k$th detector (at position $\vec{x}_k$)
for a plane wave propagating along $\hat{n}$ \cite{gwastro-PE-AlternativeArchitectures}.
We then substitute these expressions for $h_k$ into the likelihood function~\eqref{eq:lnL}
thereby generating~\cite{gwastro-PE-AlternativeArchitectures}
\begin{widetext}
\begin{align}
\ln {\cal L}(\bm{\lambda}, \theta) 
&= (D_{\rm ref}/D) \text{Re} \sum_k \sum_{\ell m}(F_k \Y{-2}_{\ell m})^* Q_{k,lm}(\bm{\lambda},t_k)\nonumber \\
&   -\frac{(D_{\rm ref}/D)^2}{4}\sum_k \sum_{\ell m \ell' m'}
\left[
{
|F_k|^2 [\Y{-2}_{\ell m}]^*\Y{-2}_{\ell'm'} U_{k,\ell m,\ell' m'}(\bm{\lambda})
}
 {
+  \text{Re} \left( F_k^2 \Y{-2}_{\ell m} \Y{-2}_{\ell'm'} V_{k,\ell m,\ell'm'} \right)
}
\right]
\label{eq:def:lnL:Decomposed}
\end{align}
\end{widetext}
where 
 where $F_k = F_{+,k} - i F_{\times,k}$ are the
complex-valued detector
response functions of the $k$th detector \cite{gwastro-PE-AlternativeArchitectures} and 
the quantities $Q,U,V$ depend on $h$ and the data as
\begin{subequations}
\label{eq:QUV}
\begin{align}
Q_{k,\ell m}(\bm{\lambda},t_k) &\equiv \qmstateproduct{h_{\ell m}(\bm{\lambda},t_k)}{d}_k \nonumber\\
&= 2 \int_{|f|>f_{\rm low}} \frac{df}{S_{n,k}(|f|)} e^{2\pi i f t_k} \tilde{h}_{\ell m}^*(\bm{\lambda};f) \tilde{d}(f)\ , \\
{ U_{k,\ell m,\ell' m'}(\bm{\lambda})}& = \qmstateproduct{h_{\ell m}}{h_{\ell'm'}}_k\ , \\
V_{k,\ell m,\ell' m'}(\bm{\lambda})& = \qmstateproduct{h_{\ell m}^*}{h_{\ell'm'}}_k  \ .
\end{align}
\end{subequations}
The likelihood can be equivalently expressed as
\begin{align}
\ln {\cal L} &= \frac{D_{\rm ref}}{D} \text{Re}[ (F Y)^\dag Q]  \nonumber \\
 & - \frac{D_{\rm ref}^2}{4 D^2} [ (FY)^\dag U FY + (FY)^TV FY]
\label{eq:lnL:MatrixForm}
\end{align}
where this symbolic expression employs an implicit index-summation convention such that all naturally paired
  indices are contracted.  The result is an array of shape (time)$\times$(extrinsic).

For each candidate set of waveform parameters $\bm \lambda$, RIFT computes the inner product arrays $U,V$ and the inner
product timeseries $Q$.  Particularly for low-mass binaries, these precomputed quantities can be costly to evaluate, as
their calculation  involves manipulating long timeseries.  By contrast, the reduced quantities $Q$ and $\ln {\cal L}$
only need to be evaluated over a very short range of times, allowing for massive reduction in computational cost; see
\cite{gwastro-PENR-RIFT-GPU} for a detailed discussion.

\subsection{Evaluating the marginalized likelihood}
\label{sec:sub:Lmarg}
Given the likelihood ${\cal L}_{\rm full}(\bm\lambda,\bm\theta)$, RIFT evaluates the marginal likelihood via an adaptive Monte Carlo
integrator:
\begin{align}
\label{eq:lnL:MonteCarlo}
{\cal L}(\bm\lambda) \simeq \frac{1}{N} \sum_k {\cal L}_{\rm full}(\bm\lambda,\theta_k) p(\bm\theta_k)/p_s(\bm\theta_k)
\end{align}
RIFT provides three core adaptive Monte Carlo techniques \cite{gwastro-RIFT-Update} to perform this and any other integral.    Two techniques -- the original
sampler \cite{gwastro-PE-AlternativeArchitectures} and a GPU-accelerated variant \cite{gwastro-RIFT-Update} -- involve an (adaptive) sampling prior $p_s$ which has
 product form, consistent with standard Cartesian adaptive integrators
 \cite{Lepage:1980dq,2021JCoPh.43910386L,book-mm-NumericalRecipies}.  
After a large block of evaluations, each one-dimensional marginal sampling prior can be updated to more
closely conform to the support of the integrand.
The other implementation uses Gaussian mixture models (GMMs) to generate sampling priors for any user-specified subset of
dimensions.
For this stage, the most recent RIFT implementation emphasize the former  GPU-accelerated adaptive-cartesian (AC)
implementation, as it quickly inferred broad posteriors.  However, for very high-amplitude sources with well-localized
intrinsic parameters, the  GMM implementation was also pertinent when computing marginal likelihoods and extrinsic
posterior distributions. 

\subsection{Likelihood interpolation and posterior distributions}
\label{sec:sub:CIP}
To estimate ${\cal L}$ from discrete samples $\lambda_\alpha,{\cal L}_\alpha$, RIFT currently deploys two unstructured
interpolation techniues:  Gaussian process regression and random forest regression.  
Given the extremely high training and evaluation cost for Gaussian processes, in practice we almost exclusively
recommend using random forest regression for production work.

Given the likelihood, fair samples from the posterior distribution are generated by the following two-step process,
described in the RIFT paper.
First, using the likelihood estimate $\hat{\cal L}_{\rm marg}$ and the same adaptive Monte Carlo integrator described above, 
we perform the Monte Carlo integral $\int d{\bm \lambda} \hat{\cal L}_{\rm marg} p({\bm \lambda})$, producing sample
locations ${\bm \lambda}_k$ and associated weights $w_k= \hat{\cal L}_{\rm marg} p(\lambda)/p_s(\lambda)$.  
Second, we make a fair draw from these weighted samples.

In typical practice, the Monte Carlo integration(s) used to generate the posterior are  performed in
parallel,  with each CIP worker instance terminating once a threshold $n_{\rm eff}$ is reached, where as described at length in
Wofford et al \cite{gwastro-RIFT-Update} RIFT uses an effective sample estimate
\begin{align}
\label{eq:neff}
n_{\rm eff} = \frac{\sum p_k}{\max \{p_k \} } 
\end{align}
expressed for convenience in terms of normalized sample probabilities $p_k = w_k/\sum_q w_q$.
The value $1/n_{\rm eff}$ is the largest discontinuous jump in the estimator $\hat{P}(<x) = \sum_k p_k \theta(x_k-x)$
for any one-dimensional cumulative probability distribution $P(<x)$ derived from the full samples.
The posterior is the union of fairly drawn samples from each CIP worker, where each worker has been configured to
fairly draw (a constant multiple of)  $n_{\rm eff}$ samples under the assumption that the $n_{\rm eff}$ threshold has
been reached  (i.e., rather than a user-supplied maximum evaluation count, to regulate runtime).
In this work, unless otherwise noted we use $n_{\rm eff}/N$ to characterize our sampling efficiency.


\subsection{Exploring the parameter space}
\label{sec:sub:explore}
For expedient convergence, RIFT has two additional methods to explore the parameter space: dithering and incremental
dimensionality, both described in detail in Sections 3.B and 3.C of Wofford et al. \cite{gwastro-RIFT-Update} and
extended with sections 4.A and 4.D of that paper.   Wofford
et al
also describes RIFT's convergence tests.
In this section, we only briefly review pertinent elements improved within this work.

After the posterior is produced and a candidate grid generated, RIFT can optionally produce a second candidate grid
derived from and supplementing the first.  In this second grid, points are generated by performing dithering on (or
randomization of) arbitrary
combinations of parameters, then rejecting unphysical combinations.  For example, the candidate points may have small
(correlated) offsets in chirp mass, $\eta$, and $\chi_{\rm eff}$  added, with offset covariance matrix set by the
covariance matrix of the input candidate grid.    
Particularly after several iterations, this dithering  can remedy a significantly-offset initial grid which misses the true
likelihood maximum.    This dithering also insures good sampling outside the boundaries of the target point.
The original RIFT paper \cite{gwastro-PENR-RIFT} only implemented correlated dithering based on sample covariance.  Later in this paper, we
describe incremental improvements to the dithering process which further improve performance.

RIFT can also employ different parameterizations at each stage.  In particular, as explained in the RIFT paper, RIFT can employ likelihood models with increasing
numbers of parameters, starting with the dominant degrees of freedom (e.g., $\mc$, $\eta$, and $\chi_{\rm eff}$ for
massive BHs) and adding in subdominant degrees of freedom in subsequent iterations.   
This approach helps address a tradeoff between cost and complexity.   For the first few iterations, RIFT needs to
identify the peak likelihood, as characterized by the dominant parameters.  Using all model parameters can be highly
counterproductive, as fits with all degrees of freedom require overwhelming numbers of evaluations  $\lambda_k, {\cal
  L}_k$ in order to avoid overfitting/under-resolving.  (With too few evaluations and several irrelevant parameters
included, our random forest or  Gaussian process fits behave pathologically, albeit in different ways.)    By reducing the number of poorly-constrained parameters early
on, we can employ far fewer points early on.
%
The appropriate dimensional hierarchy depends on the physics involved (e.g., configurations with high
mass; BHNS with strong precession; NS-NS binaries with tides; et cetera) but is well-motivated from simple Fisher matrix
arguments.
Specifically, the component masses and a measure of aligned binary spin (e.g., $\chi_{\rm eff}$) approximately
characterize the dominant degrees of freedom for nearly-nonprecessing binaries, particularly when organized as the chirp
mass $\mc$ and symmetric mass ratio $\eta$.  As most observed binaries exhibit
nearly no precession, these variables form a natural set to adopt for the first iterations.  For binaries not exhibiting
strong precession-induced modulations on our line of sight,  transverse and other
spin degrees of freedom have a subdominant impact on the marginal likelihood, so we can add these incrementally, after
obtaining a converged estimate for the behavior for nonprecessing degrees of freedom.

\subsection{Source-dependent pipeline tuning}
Relying on an external queuing system for key control features and feedback, RIFT currently does not adaptively
reconfigure the effort needed during an analysis.  Instead, all operating point choices such as the number of likelihood
grid points evaluated by ILE and the number of CIP workers used per iteration must be pre-specified when the pipeline is
built.  The number of ILE evaluations is controlled by the problem dimension based on past experience with
modest-amplitude gravitational wave sources with different source amplitudes and parameters, as outlined in previous work.  Starting
with the 0.0.15 releases, RIFT includes a  mechanism (\texttt{cip-explode-jobs-auto}) to also pre-select the number of
CIP workers used to generate each posterior.  By default, we choose $N_{\rm
  cip}$ to be roughly $2+2(\rho/15)^{1.3}/q$, where we scale with the SNR $\rho$
and the mass ratio $q$, with more workers requested if matter is present or if
the end-user requests this simple estimate to be scaled up by a
multiplicative factor.

\subsection{Standard validation infrastructure}
\label{sec:sub:pp}

Though RIFT's source tree includes mulitple internal validation tests, typical users will be most interested in the
classic probability-probability (PP) test \cite{mm-stats-PP,gwastro-skyloc-Sidery2013}.
For RIFT, we formulate a PP test as a direct constraint on the cumulative distribution function.  Specifically, using
RIFT on each source $k$ in a fair sample of sources drawn from a specific (intrinsic plus extrinsic) prior with true parameters $\mathbf{\lambda}_k$, we estimate
the fraction of the posterior distributions which is below the true source value $\lambda_{k,\alpha}$   [$\hat{P}_{k,\alpha}(<\lambda_{k,\alpha})$] for each intrinsic parameter $\alpha$.  After reindexing the sources so $\hat{P}_{k,\alpha}(\lambda_{k,\alpha})$ increases with $k$ for some fixed $\alpha$,  a
plot of $k/N$ versus $\hat{P}_k(\lambda_{k,\alpha})$ for both mass parameters can be compared with the expected result
($P(<p)=p$) and binomial uncertainty interval.

RIFT's source tree provides two types of automation to facilitate PP tests.  With the first level of automation, we
employ a simplified version of the RIFT algorithm, without using the complexities (e.g., subdags and iteration to
convergence, as described in Wofford et al.).  Though calling the same underlying codes, this simplified algorithm does
not exploit all the flexibility developed in Wofford et al for flexible architectures and iteration to convergence, nor
does it produce extrinsic posterior samples. 
That said, this lightweight approach is extremely well suited to simplified tests, particularly involving nonprecessing
waveform development and model systematics. 
For more complex scenarios such as binaries which may precess,  we now provide automation that uses the ``standard'' mode
of operation, generating synthetic sources and then calling the full RIFT pipeline.  This infrastructure allows us to
test RIFT using precisely the same inputs and modes of operation as used in single-source production-quality inference. 

\subsection{External interface: Waveforms}
\label{sec:sub:external:Waveforms}

Building on  several recent investigations, this work reports on  enhancements to RIFT to improve its access  interfaces or otherwise updated its approach to a few notable frameworks.
This paper reports on providing RIFT access to the new lalsuite waveform interface (\texttt{gwsignal})  \cite{ligo-gwsignal}.
This interface supports the latest waveform developments, notably including TEBOResumS-Dali \cite{2024PhRvD.110h4001N}, SEOBNRv5PHM
\cite{2023arXiv230318046R,2023arXiv230318203M}, SEOBNRv4EHM \cite{2022PhRvD.105d4035R},   SEOBNRv5EHM \cite{gwastro-seob-v5EHM-2024}. 
A recent study updated our direct TEOBResumS interface to enable access to unbound (hyperbolic) orbits
(Henshaw et al in prep \cite{gwastro-hyperbolic-Henshaw2025}).
Another study  expanded our infrastructure to allow for both eccentricity and the mean anomaly \cite{gwastro-pe-eccentric-Wagner2024}.
For the old lalsuite ChooseFDWaveform interface, we are improving conditioning and corrected erroneously-interpreted
extrinsic Euler angles, as described at greater length in Appendix \ref{ap:hlm_conventions}.

Additionally, RIFT now provides mechanisms for users to efficiently pass extra arguments to the lalsimulation and gwsignal waveform
generators.  For the lalsuite interface, we provide a command-line argument for ILE
(\texttt{internal-waveform-extra-lalsuite-args}) which allows the user to specify a python-formatted dictionary of key-value
pairs, passed to the low-level lalsimulation interface using the \texttt{SimInspiralWaveformParamsInsertNAME} template for
various choices for \texttt{NAME}.  Among other capabilities, this interface allows the user to access different versions of
lalsimulation waveforms, such as the IMRPhenomXPHM \cite{Pratten_2021} family, which has undergone a number of point release version updates
as this paper was developed. To access these different versions, the user can for example specify \texttt{ \{
  PhenomXHMReleaseVersion: 122022 \} } (the current default)  or similarly use \texttt{122019} for a previous release.
More information about different options for the PhenomXHM models is available in their source code \cite{lalsuite}.
For the \texttt{gwsignal}, we provide a different command-line argument (\texttt{internal-waveform-extra-kwargs}) for the
user to provide a python dictionary, fed directly to the gwsignal waveform generator.  For example, for the SEOBNR model family, this
mechanism can be used to pass the \texttt{lmax\_nyquist} argument, which controls the maximum $L$ used in a test to assess
whether aliasing will occur at the end of the signal.  (The $(L,L)$ mode is roughly $L$ times higher frequency than the
$(2,1)$ mode; thus, a waveform generating many higher-order modes may have some of its content undersampled and
therefore alias during or after the merger phase at the end of the signal.  Using a small value for this parameter allows these models to be used with much
lower sampling rates, as needed for compatibility with non-RIFT PE frameworks and operating choices.)
These arguments can be set within RIFT's configuration file, or with the \texttt{asimov} interface described later.

\subsection{External interface: Calibration marginalization}
\label{sec:sub:external:calmarg}

RIFT analyses for GWTC-3 were performed with a postprocessing step to marginalize over detector calibration
uncertainty.  This postprocessing involved generating a weight $W_\alpha$ for each sample $x_\alpha$ derived from the
RIFT-estimated gravitational wave strain $h(x_\alpha)$ associated with that sample and a fixed number $N_{\rm cal}$
realizations of the detector calibration uncertainty, then performing rejection sampling.  RIFT used an infrastructure
originally derived for a different study \cite{2020PhRvD.102l2004P}, ingesting calibration uncertainties and performing
the reweighting calculation with the external \texttt{bilby} library \cite{2019ApJS..241...27A}.   Specifically, using the gravitational wave
strain provided by RIFT (either resumming the raw $h_{lm}(t)$ directly or by calling $h(t)$, to allow for different
conventions in these two interfaces), the \texttt{calibration\_reweighting} script uses the built-in bilby calibration
reweighting code for the \texttt{bilby.gw.GravitationalWaveTransient} likelihood to compute the relative likelihoods
$1/W_\alpha = L(x_\alpha| C_{ref})/\int dC L(x_\alpha|C)p(C)$, where $C$ denotes a calibration realization and $p(C)$
the prior over calibration realizations.
To facilitate automated self-contained end-to-end investigations including this implementation of calibration marginalization, we have
incorporated an updated version of this script into the RIFT source code and data analysis pipeline.

\subsection{External user-provided likelihods and priors in ILE, CIP}
\label{sec:sub:external_lnL}
Though not previously described in a publication, RIFT has provided a simple mechanism since version 0.0.15.8 allowing
users to supplement the GW
likelihood ${\cal L}$ with an external likelihood ${\cal L}_{\rm ext}$ (e.g., to perform inference using ${\cal L}\times {\cal L}_{\rm ext}$), without needing to access or modify RIFT itself.  In previous publications, these extensions have been
used for example to impose an observationally-motivated inclination prior for GW170817 \cite{gwastro-RIFT-hyperpipe-EOS-Atul,gwastro-ns-eos-Vilkha2024}.  
Specifically, both the extrinsic (ILE) and intrinsic (CIP)
integration codes provide the end user with three arguments:
\texttt{--supplementary-likelihood-factor-code}, to provide the name of a module to import;
\texttt{--supplementary-likelihood-factor-function}, to identify the specific function in this module to be called; and
\texttt{--supplementary-likelihood-factor-ini}, to allow the end user to supply a (configparser-compatible)
configuration file, which RIFT will parse and pass to a suitably-named program in the user-provided module.
For ILE to work with this likelihood on a GPU, this supplementary likelihood may need to conform to all necessary
conventions for typing and data handling, as illustrated in our code.
While our external likelihood mechanism has numerous science  applications in  multimessenger inference, in this work we
will use this interface to facilitate controlled end-to-end tests of new integrators within our existing tools.

Additionally (albeit somewhat redundantly in principle), users can override any of the priors employed in inference
within CIP, using a similar framework.  Specifically, the user provides \texttt{--supplementary prior code}, which must
have two dictionary members \texttt{prior\_pdf} and \texttt{param\_ranges}, each of which directly overrides the
corresponding default entry inside CIP.  Because CIP does not currently use GPU aceleration, these prior routines do not
need to access any GPUs either. 

\subsection{Comments}
\label{sec:sub:comments}

RIFT's full potential remains untapped, as we conservatively relied on brute force.  For context, the most
parsimonious new prototype strategies like Dingo-IS \cite{gw-astro-mergers-dingo-part2} might  produce one
independent sample for every  $10$ likelihood evaluations, or an efficiency $\epsilon\simeq 0.1$, only needing $O(10^5)$
likelihood evaluations to produce $10^4$ independent samples; 
the best new samplers compatible with contemporary approaches might rquire $10^7$ evaluations ($\epsilon \simeq 10^{-3}$)
\cite{2021PhRvD.103j3006W,gwastro-mergers-nessaiv2-2022};
while conventional strategies often have $\epsilon\simeq O( 10^{-4})$  \cite{2021MNRAS.507.2037A,gwastro-mergers-nessaiv2-2022}, requiring $O(10^8)$
likelihood evaluations.
By contrast, while a production-quality unsupervised RIFT analysis might only simulate radiation from  $O(2\times 10^4)$ distinct binaries, the
likelihood will be evaluated $O(10^{11}-10^{12})$ times, in part  because every Monte Carlo integral at every stage of
our analysis is performed completely independently, not leveraging any prior experience!  [Despite the many evaluations
  needed, where we refer to an evaluation as everything including waveform
  generation \textit{and} likelihood calculation, RIFT has remained 
  runtime-competitive  with other production-ready methods because of our extremely low-cost likelihood and highly
  parallelizable architecture \cite{gwastro-PENR-RIFT,gwastro-PENR-RIFT-GPU,gwastro-RIFT-Update}. RIFT also explores the
whole space $\bm{\lambda},\bm{\theta}$, not relying on pretraining a sampling oracle like Dingo-IS.]

A fortuitous outcome of RIFT's structural inefficiency has been automatic redundancy: for our posterior generation code,
even within a single event analysis, similar
calculations are naturally  performed with very different settings (i.e., coordinates, priors, and sampler settings). As
discussed in Section \ref{ap:checks}, 
these built-in self-consistency checks
provide critical opportunities to identify and diagnose subtle issues that can arise in parameter
inference.

\section{RIFT pre-O4 testing}
\label{sec:tests_old_version}

During preparations for O4, we examined the performance of the standard settings recommended in the preceding study
\cite{gwastro-RIFT-Update}: ILE operated with adaptive cartesian (AC) integration, limited extrinsic adaptation (i.e.,
in the sky position and inclination), batched in modest-size collections for efficiency; and CIP operated with GMM using
transverse spin coordinates for most iterations.  
For most detectable sources, this default configuration performed very well.  For louder sources (SNR greater than about
30), where almost all extrinsic parameters can be well-localized, we unsurprisingly found that our extrinsic integration (ILE) was much more efficient
when we allowed for adaptation in all extrinsic parameters, particularly for high-mass ratio sources whose long
inspirals enabled extremely precise extrinsic parameter measurements.

\begin{figure*}
  \includegraphics[width=\columnwidth]{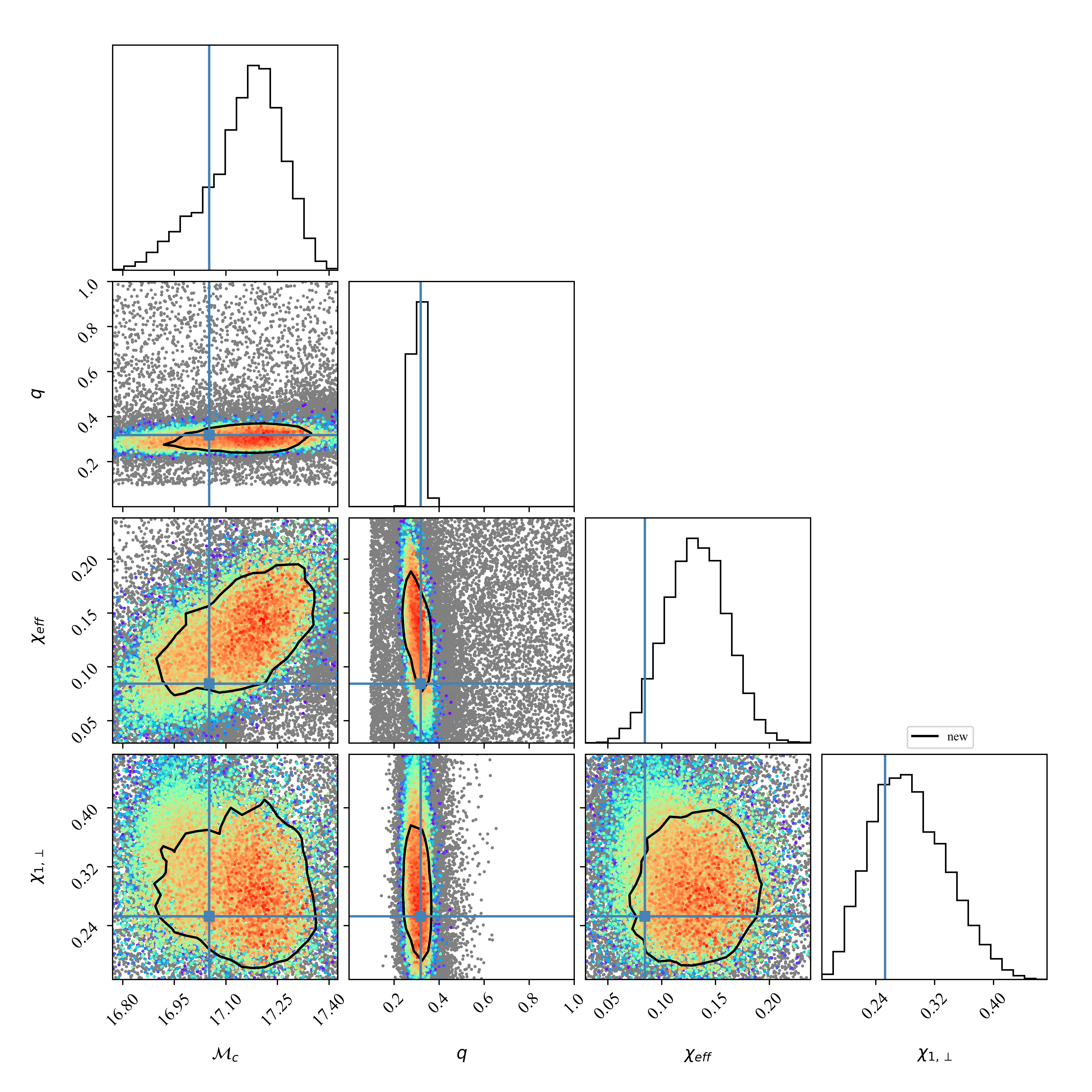}
  \includegraphics[width=\columnwidth]{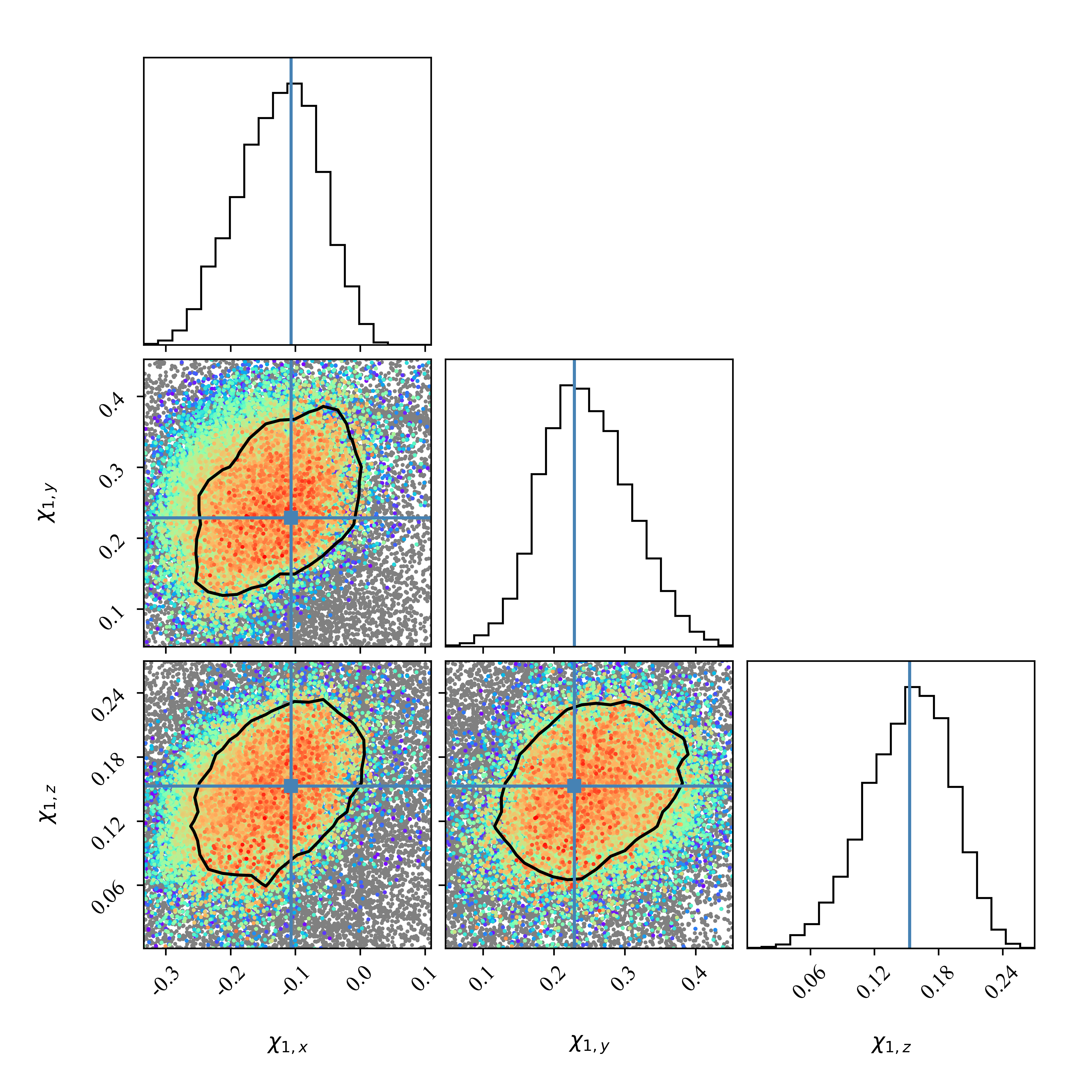}
\caption{\label{fig:fiducial_basic} \textbf{Inference of a fidicual binary black hole}: Panels show corner plots with
  90\% credible interval posterior quantiles, for inference of a synthetic binary black hole signal generated and
  interepreted with IMRPhenomXPHM. The color
  scale shows the likelihood range, over a dynamic range $\Delta \ln {\cal L}\le 15$; gray points indicate likelihood
  evaluations below this range.  The panels show marginal distributions in chirp mass $\mc$, mass ratio $q=m_2/m_1$,
  inspiral effective spin $\chi_{\rm eff}$, and the magnitude of the transverse dimensionless spin of the primary
  $\chi_{1,\perp} = \sqrt{\chi_{1,x}^2+\chi_{1,y}^2}$. 
}
\end{figure*}

\begin{figure}
  \includegraphics[width=\columnwidth]{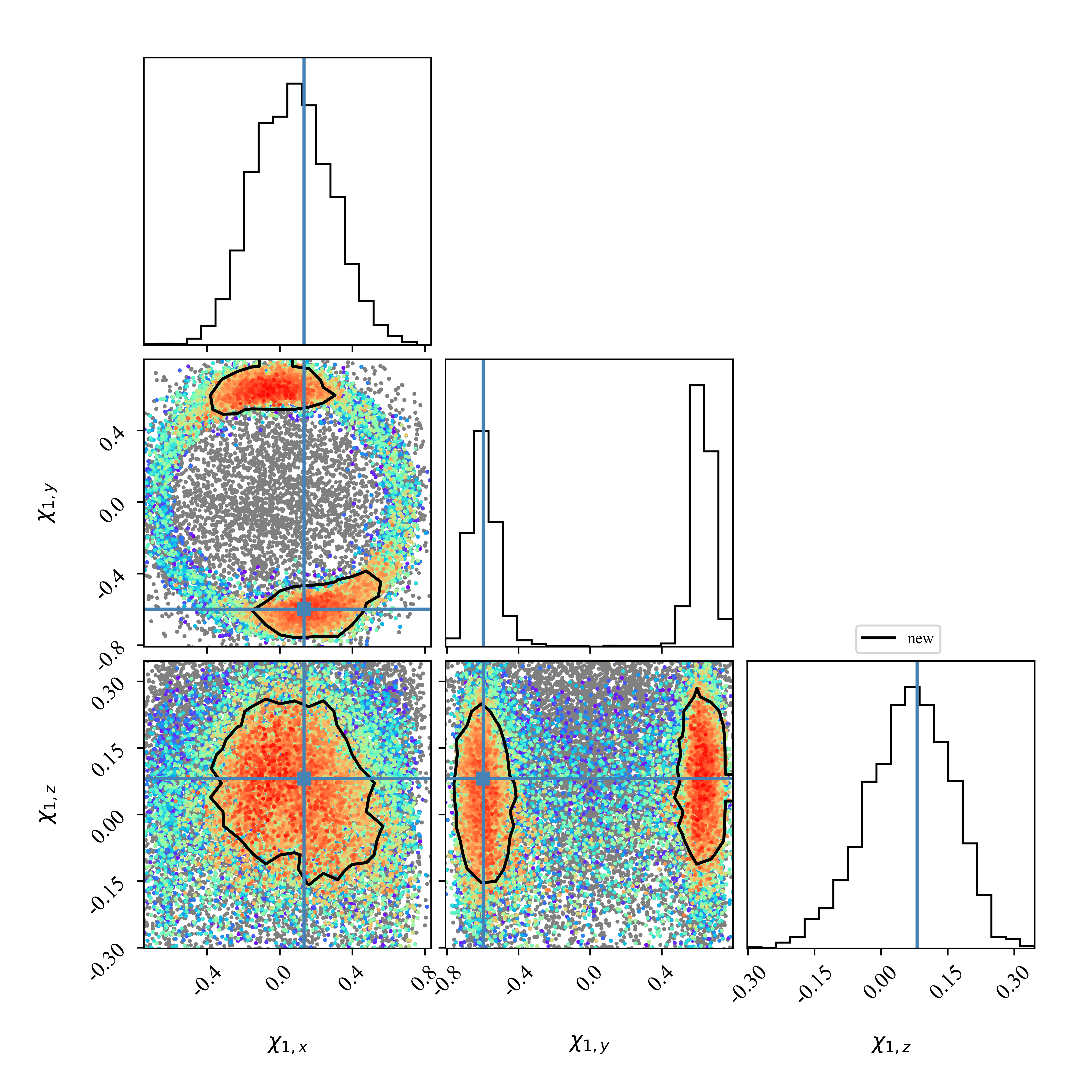}
\caption{\label{fig:weak:spin_lobes}\textbf{Example of multi-modal intrinsic parameters for spins}: Using a different
  synthetic event, panel shows corner
  plot with 90\% credible interval posterior quantiles, similar to Figure  \ref{fig:fiducial_basic} but illustrating
  cartesian components of the dimensionless primary spin. For this event, the posterior distribution for primary spin
  has two modes. }
\end{figure}

\subsection{Weaknesses of  the pre-O4 configuration}
Despite robust performance across a wide range of mass, mass ratio, and spin scales (e.g. long-duration $q\simeq 20$ binaries with
large transverse spin), we identified a few anecdotal examples where RIFT's contemporary performance should be improved.
First and foremost, we found sources with highly localized extrinsic parameters (e.g., edge-on binaries with large $q$)
required special attention at modest and even low SNR: like their high-SNR counterparts, they benefitted from adaptation
in more extrinsic parameters.    Figure \ref{fig:fiducial_basic} shows an example  of the parameters and interpretation
of such a configuration; the analysis settings used are described in Section \ref{sec:results}. 
Second, we found that binaries with significant transverse spin and high mass ratio have either a ring,  bimodal, or
rarely unimodal shape in the  $\mathbf{\chi}_1$ sphere: either a ring  around the $\hat{L}$ axis, a bimodal pair with
who lobes offset by $\pi$ around the axis, or rarely a single dominant lobe a the highest amplitudes.  Figure
\ref{fig:weak:spin_lobes} illustrates examples of these configurations and the posterior distribution of their
parameters; again, analysis settings are described in Section \ref{sec:results}. 
While both our transverse and polar spin coordinates are well-adapted to these  ring shapes,  at times our architectures
and samplers simply insufficiently well explored these regions.    For example, the default configuration operates the
first few iterations with a fixed prior on transverse spins, strongly disfavoring large transverse spins.  This
configuration was very inefficient at discovering large transverse spins in high-mass-ratio systems, whose transverse
spin posteriors are very well localized.
Third, we found that loud or extremely localized signals could have large error in ${\cal L}$, due to how we
operated ILE (e.g., the choice of integrator; a fixed and too low  maximum number of evaluations; and, critically, the
choice to sample uniformly over several extrinsic variables, not adaptively).  These large
integration errors either led to  no output at all or, perhaps worse, output with so large errors that downstream stages
simply ignored them.  As a result, while the pipeline operated, its output was based on a handful of high-uncertainty
evaluations at each iteration, producing unreliable posteriors.   In other words, the pipeline failed to identify
consistently poor ILE outcomes and adapt its ILE integration settings accordingly.
Finally, we realized that the interface to IMRPhenomXPHM adopted a frame convention inconsistent with all other
interfaces and  our operating assumptions: the $h_{lm}(t)$ were specified in a frame aligned with the total rather than
orbital angular momentum; see Appendix \ref{ap:hlm_conventions} for details.



\begin{figure}
  \includegraphics[width=\columnwidth]{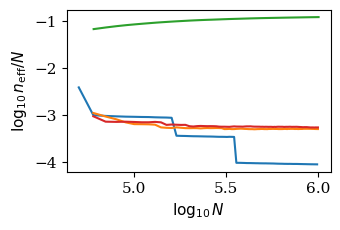}
\caption{\label{fig:n_eff_demo}Sampling efficiency versus $N$ for the Rosenbrock problem, showing the AV integrator
  (green), the GMM integrator (blue), the AC integrator (orange), and the original
  default integrator (red). $n_{\rm eff}$ is the number of effective
  sample points to calculate before the integration will terminate and $N$ is
  the total number of extrinsic points to evaluate at, including waveform
  generation and marginalized likelihood evaluation. }
\end{figure}

For all our analyses, the low-level integrators used in ILE and CIP had extremely low sampling efficiency, increasing
our computational cost and runtime.  Figure \ref{fig:n_eff_demo} illustrates the sampling efficiency of the previously-standard
integrators using a standard test problem (the Rosenbrock likelihood) distributed with RIFT's source code.  For
comparison, this figure also illustrates the corresponding sampling efficiency of the new AV integrator described in Section \ref{sec:improve_int}.

\subsection{Stopgaps and minor fixes for the pre-O4 configuration}
Before deploying the updates described later in this paper, we provided  a few stopgap suggestions  to mitigate the
issues seen above for users of the previous edition (version 0.0.15.11).
First and foremost, we updated our interface to IMRPhenomXPHM to better align the waveform with our assumptions; see  Appendix \ref{ap:hlm_conventions}.
Second, we recommended choosing RIFT's settings based on prior experience with a specific analysis.  Specifically, events with strong
suspicion of possessing large transverse spin should be operated with conventional spherical polar coordinates (and
sampling priors, early on), to better explore the large transverse spin space.  Likewise, events with strong suspicion
of being highly localized in extrinsic parameters should be operated with all parameters being simultaneously adapted.


\subsection{Followup on selected RIFT O3 results}

Since the original O3 LVK analyses and in the context of other investigations,
 \cite{2024CQGra..41a5005G}  performed followup investigations of a few O3 events 
where the original LVK analyses done with RIFT suggested substantial systematics.   In their reanalyses of two such
high-detector-frame-mass events (GW190527 in GWTC 2.1 and GW190719 in GWTC-2),
they found their reanalysis with parallel-bilby \cite{Smith_2020} and SEOBNRv5PHM produced
results comparable to the other results presented in that work, with modest disagreement relative to the
RIFT/SEOBNRv4PHM analyses.  [The authors' reanalysis of three other similarly identified events did not produce notable
  changes.]

\begin{figure*}
\includegraphics[width=0.3\textwidth]{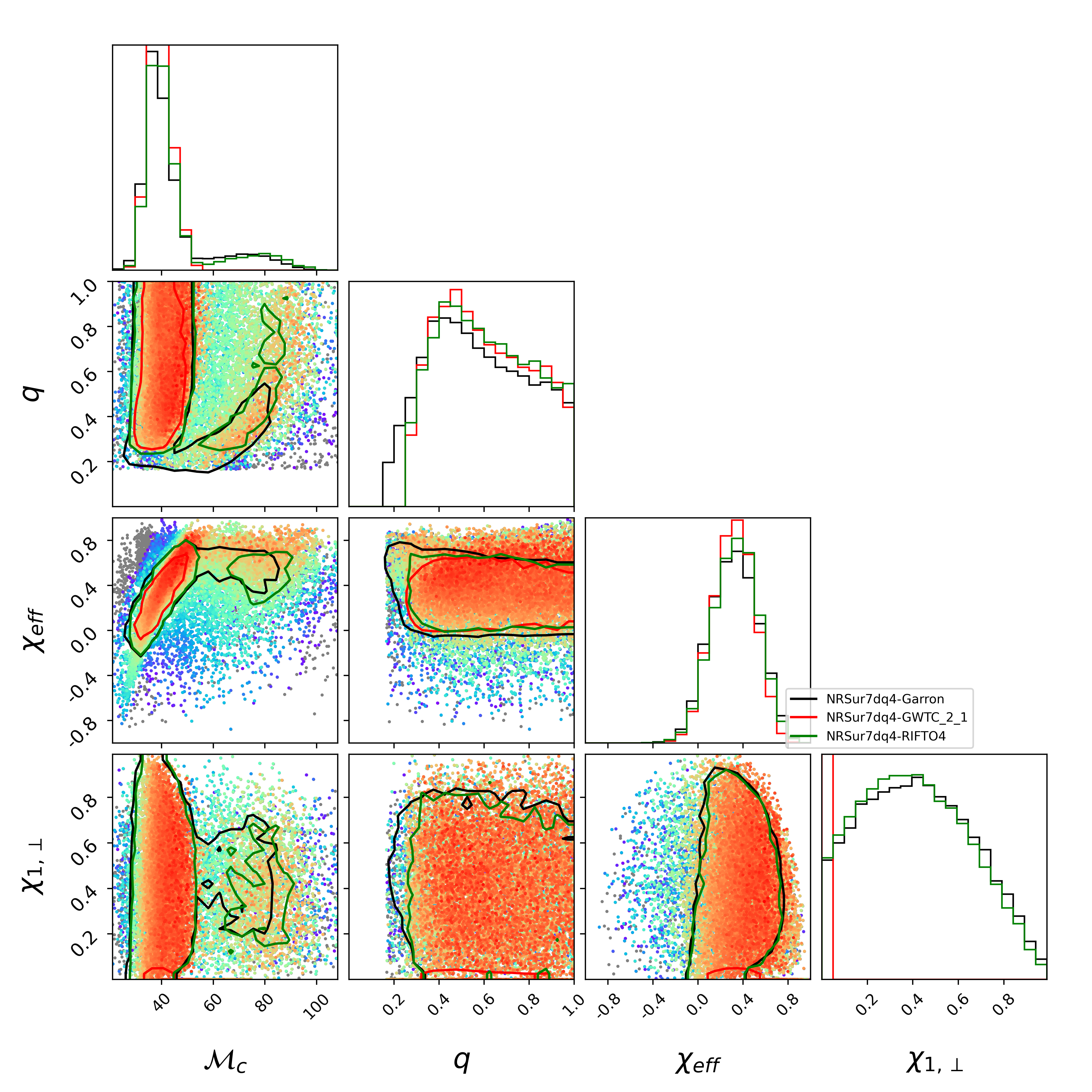}
\includegraphics[width=0.3\textwidth]{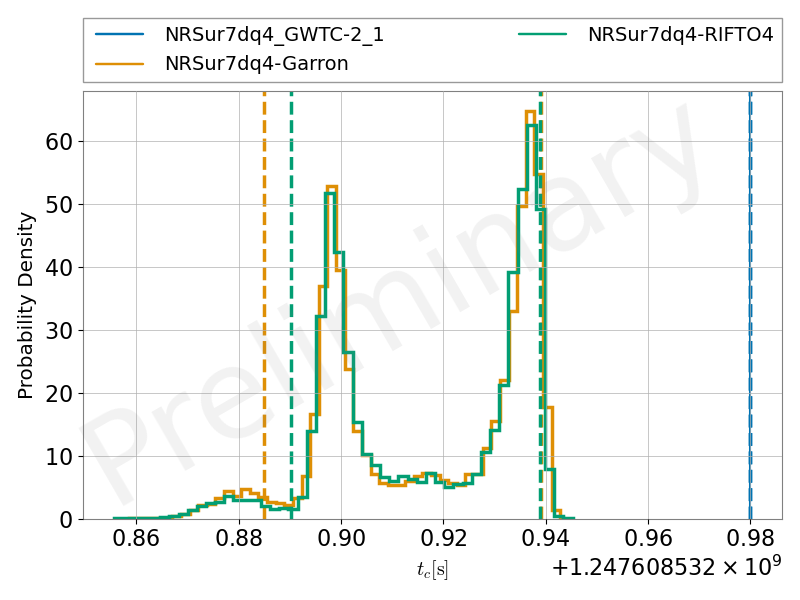}
\includegraphics[width=0.3\textwidth]{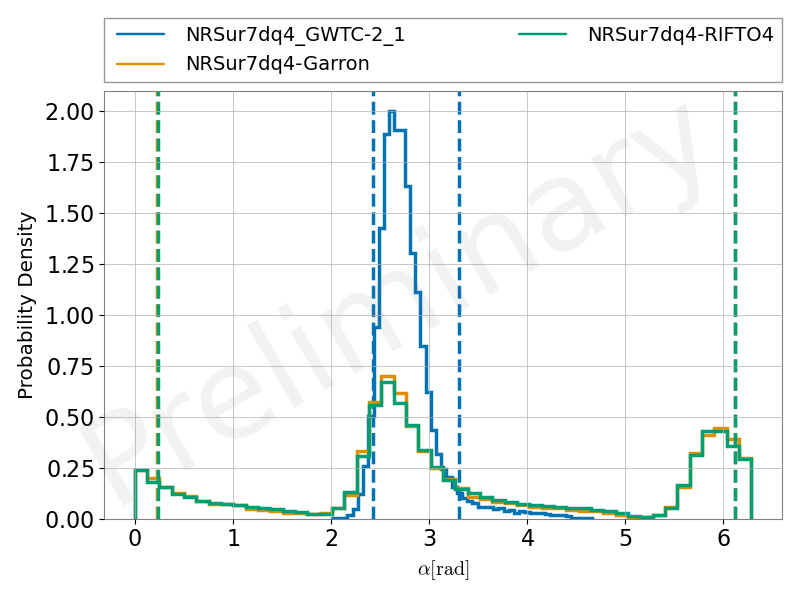}
\caption{\label{GW190719_rerun}
\emph{Left panel}:Corner plot showing comparisons of GW190719 data with NRSur7dq4. The red and green analyses are
performed with RIFT: the black analysis is the fiducial LVK result provided in GWTC-2.1, while the green curve shows our
revised analysis with an updated analysis time window.  For comparison, the black curve shows the analysis reported in
\cite{2024CQGra..41a5005G}  for this event.
\emph{Center panel}: One-dimensional marginal time distribution.  The GWTC-2.1 edition of RIFT (shown here in blue) only provides a single
number: the center of the analysis window.  The GWTC-2.1 analysis central time window only extends down to $\simeq 0.93$
on this axis and thus does not include most of the one-dimensional marginal distribution.
\emph{Right panel}: One-dimensional marginal distribution of right ascension.  Due to its truncated time window, the
GWTC-2 analysis produces a distorted skymap}
\end{figure*}

At least some of these regrettable circumstances reflect user error when establishing settings.  The analysis of
GW190719 inadvertently adopted a central event time too far offset from the recovered event time, so some of the
posterior was excised as out of the inference coincidence window.  [Like all parameter inference codes, RIFT employs a
  finite geocentric time window to select data for analysis.  The default RIFT time coincidence window in GWTC-2 was
  $\pm 50\unit{ms}$, centered on a target trigger time.]  Figure \ref{GW190719_rerun} shows the results of 
analyses of GW170719 with the current version of RIFT with NRSur7dq4 \cite{Varma_2019}, compared with the previously reported
GWTC-2 results.    The limited time coincidence window excises a substantial fraction of the skymap and distorts the
intrinsic posterior.  Our revised analysis agrees with the results provided by  \cite{2024CQGra..41a5005G}, keeping in
mind that we adopt the same mass ratio prior lower bound as GWTC-2
\cite{LIGO-O3-O3a-catalog} in the results presented here.

\begin{figure}
\includegraphics[scale=0.3]{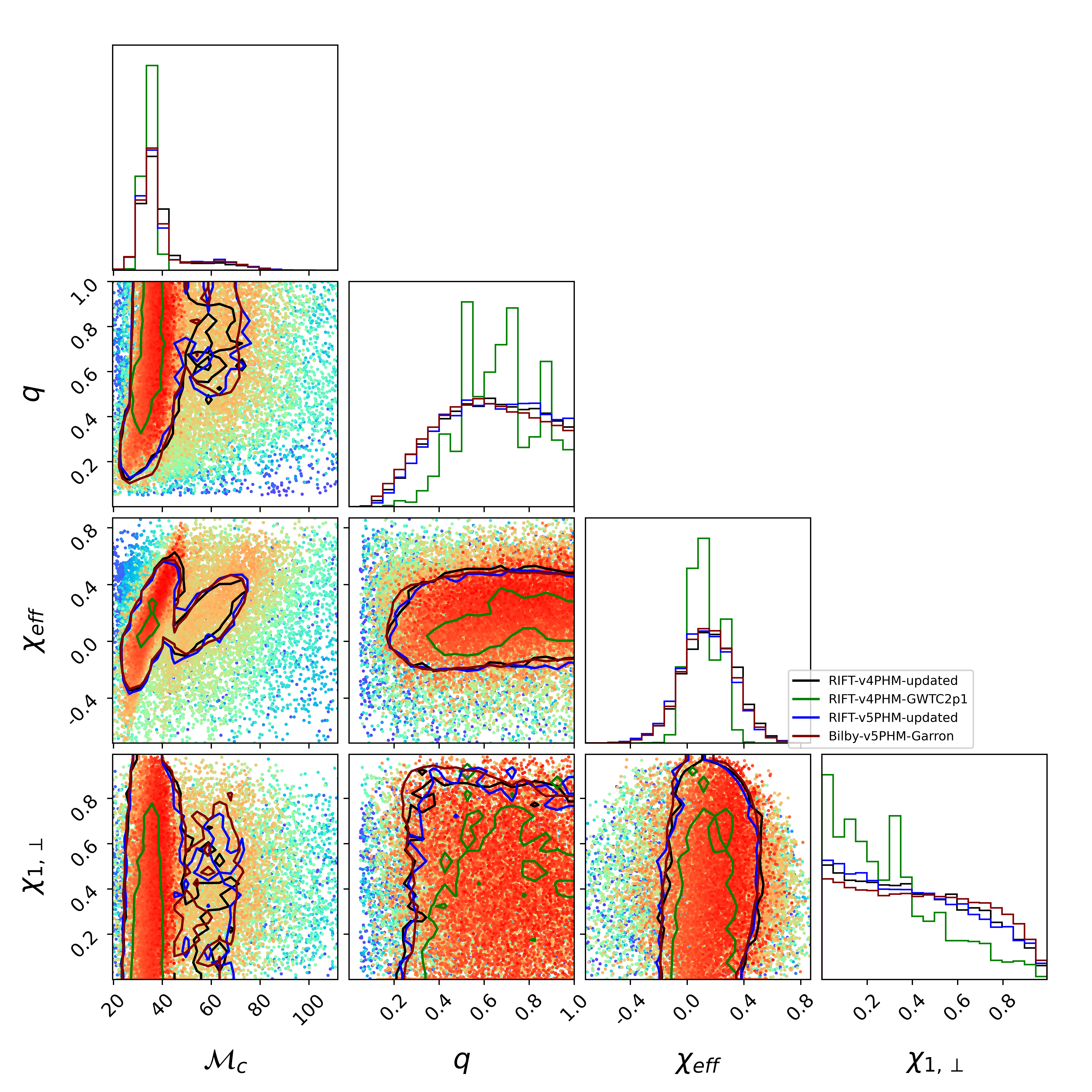}
\caption{\label{GW190527_rerun}
Corner plot showing comparisons of GW190527 data with SEOBNRv4PHM.  In this figure, the green curve
shows the reference RIFT analysis performed for GWTC-2.1, which adopts a high starting frequency (and employs
calibration marginalization, thus decimating its samples).  The black and red curves present reanalyses performed with
RIFT in this work and with bilby in  \cite{2024CQGra..41a5005G}; both agree.   For context and to demonstrate waveform
systematics are small, the blue curve also reports an analysis with SEOBNRv5PHM.}
\end{figure}

For GW190527, the lowering the starting frequency for the waveform template for SEOBNRv4PHM resulted in better agreement with  \citet{2024CQGra..41a5005G}, as shown in Figure \ref{GW190527_rerun}. 


\section{Other published enhancements}
\label{sec:previous_work}
Before describing the ewn contributions of this work, we first briefly summarize two key targeted  enhancements to RIFT
\cite{gwastro-mergers-rift_asimov_O3-Fernando2024,gwastro-RIFT-hyperpipe-EOS-Atul} that have been  deployed and demonstrated
since the last major RIFT methods paper  \cite{gwastro-RIFT-Update}.  We defer a discussion of nonparametric EOS
inference to \cite{gwastro-ns-eos-Vilkha2024} and other work in preparation. 

\subsection{Automation infrastructure}
RIFT now maintains a plugin interface to  \texttt{asimov} \cite{gwastro-mergers-rift_asimov_O3-Fernando2024}.  This package provides a code-neutral framework to encode the
event-specific settings used when interpreting individual events, as well a mechanism to efficiently perform these
inferences with multiple analysis settings.  This automation infrastructure therefore efficiently allows end users to
reproduce RIFT-based analyses of any event with RIFT, with other algorithms (e.g., \texttt{bilby}), and with other
choices for event-specific assumptions.
Since that study, we have further refined our interface, notably adding the ability to automatically use ``bootstrap''
results from previous or other analyses as an initial grid.

Due to RIFT's relatively low cost, RIFT can conveniently employ high-cost multimodal time-domain waveforms.  As a
concrete example, previous work using this automation  analyzed many events using costly models including SEOBNRv4PHM
and SEOBNRv5PHM \cite{gwastro-mergers-rift_asimov_O3-Fernando2024}.  Later in this work (e.g., in Section \ref{sec;sub:ex:ecc}),
we analyze several of those events with RIFT/asimov to further demonstrate this capacity.

\subsection{RIFT for generic inference}
The original RIFT implementation, including  data formats and multiple interface elements, are structured  specifically for coalescing
compact binaries: while highly modular in structure, it has highly restrictive data formats, interfaces, and tools.   To facilitate its re-use across a wide range of Bayesian inference  applications where individual RIFT
elements have already been employed (e.g., simulation-based
inference of binary evolution \cite{popsyn-gwastro-STInterpFinal-Vera2023,gwastro-wd-DelfaveroCosmic-2024};
estimating properties of ejected matter responsible for kilonova \cite{gwastro-nsnuc-kilonova-Ristic-JointGWEM-2025,gwastro-kilonova-YingleiNN2024}; or inference of the nuclear equation of state  \cite{Ristic22,PhysRevResearch.5.013168,gwastro-nsnuc-kilonova-Ristic-JointGWEM-2025}, a recent study \cite{gwastro-RIFT-hyperpipe-EOS-Atul}
 re-implemented the RIFT algorithm with a 
generic interface: the ``hyperpipeline''.   Both the original and re-implemented RIFT are distributed together, rely on common components, and
can interoperate (e.g., for hierarchical inference of the nuclear equation of state.
Following \cite{gwastro-RIFT-hyperpipe-EOS-Atul}., we briefly summarize this re-implementation, highlighting the opportunities it affords.

\begin{figure}
  \includegraphics[width=\columnwidth]{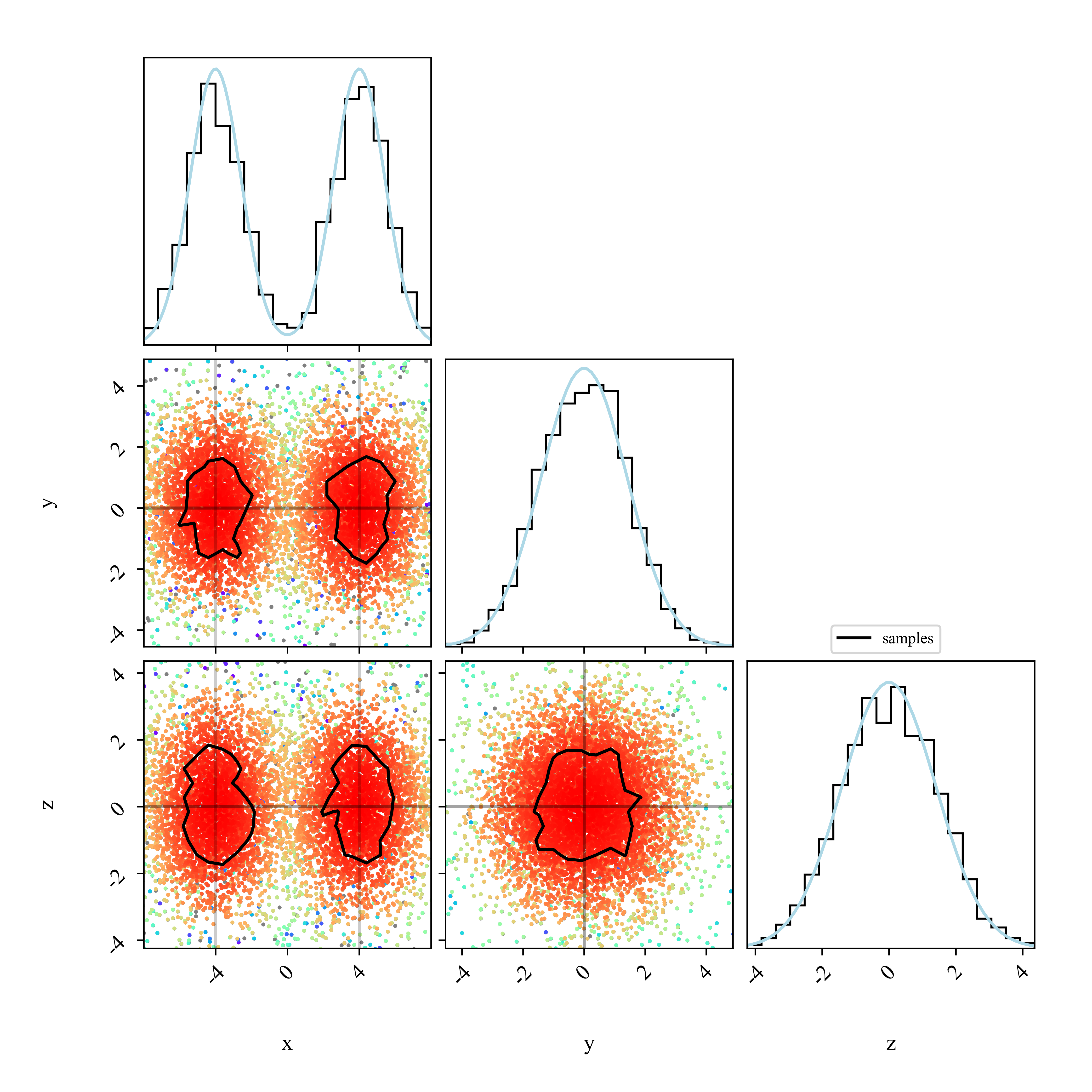}
  \label{fig:hyperipipe_demo:two_gaussians}
  \caption{\textbf{Hyperpipeline demonstration} showing recovery of a superposition of two multivariate gaussians with
    one-dimensional standard deviations $\sqrt{2}$, centered at $(\pm 4,0,0)$ respectively. The black lines in the
    corner plot correspond to inferences derived from the hyperpipeline; colors show the log-likelihoods reported by the
    external program, using our conventional scale; and the solid blue lines on the diagonal show the exact
    one-dimensional marginal PDFs.
    }
\end{figure}

Like the original RIFT, this re-implementation uses adaptive placement for parameter $\lambda_k$, using user-supplied
``worker'' jobs responsible for ingesting the parameters $\lambda_k$ and returning (factors in) the overall likelihood
${\cal L}(\lambda)$.   Also like the original RIFT, each iteration we construct a fit to $\ln {\cal L}$ and use Monte
Carlo integration to both estimate the posterior distribution and select new parameters for further investigation.  The
hyperpipeline has access to all the same low-level tools for interpolation, Monte Carlo integration, and convergence
testing.  
 Unlike the original RIFT, however, this re-implementation allows for a multi-factor likelihood
\begin{align}
\label{eq:L:hyperpipe}
{\cal L}(\lambda_k|\Lambda) = p(\Lambda) \prod_A \prod_{\alpha_A} \ell_{\alpha_A,A}(\lambda_k|\Lambda)
\end{align}
where for convenience the factors $\ell_{\lambda_A,A}$ in the likelihood are grouped into classes $A$ containing similar
kinds of observations (e.g., gravitational wave observations versus pulsar observations).  The end user supplies the
executable needed to evaluate  $\ell_{\alpha_A,A}(\lambda,\Lambda)$, an optional file of metadata unique to object
$\alpha_A$ needed to localize the evaluation for that specific object (e.g., a table of marginalized likelihoods derived
from a specific GW observations),  and a python routine to evaluate $p(\Lambda)$, based
on a common data format.    As with the original RIFT, the
new hyperpipeline implementation manages the entire iterative process, including task  scheduling needed to evaluate
$\ell_{\alpha_A,A}$ efficiently; construction of ${\cal L}(\lambda_k)$ for each iteration; and additional features to
inject randomness (``puffball'') and assess convergence needed for robust operation.
As a concrete toy model demonstration, reproducible using instructions disseminated with this paper,  Figure
\ref{fig:hyperipipe_demo:two_gaussians} shows  hyperpipeline results applied to a program which reports as log
likelihood the log-pdf from superposition of two multivariate Gaussians.  

\section{Technical advances: Improved integration}
\label{sec:improve_int}



\subsection{Adaptive volume integration}
Motivated by nested sampling, a recent study  \cite{gw-astro-mergers-VTiwariPE} introduced an
adaptively-refined volume sampler: VARAHA.  In this approach, the integration volume is subdivided into hypercubes; sampled;
hypercubes containing negligible amounts of probability are neglected; and the hypercube grid is reassessed.   The selected hypercubes will mimic whatever complex structure is present in the posterior,
allowing this  integrator to capture extremely complex features with high fidelity and with very high sampling efficiency.  

The adaptive volume approach uses a uniform-in-coordinate sampling prior $p_s$ over the unit coordiante hypercube.  For
simplicity, in this section we will describe the integration of a generic non-negative-definite function $L(x)$ over a
volume $V_0=\int dx$.   The coordinate volume and total ``probability'' inside any specific likelihood threshold $L_*$ are
given by 
\begin{align}
V(L_*)& = \int \Theta(L(x)-L_*) dx \\
P(L_*) &=  \frac{ \int \Theta(L(x)-L_*)  L(x) dx/V_0
}{
\int  L(x) dx/V_0
}
\end{align}
The algorithm will try to target a certain enclosed probability $P_{thr}$, adaptively selecting subvolumes and a
likelihood threshold $L_{thr}$, using a specific targeted number of trial samples $N$ per iteration.
At any iteration $g$, the integrator has a specific number of bins $n_{bins}$ per dimension, producing a large
potentially-accessible number of cells (i.e., $n_{bins}^d$).   Only  some
small  subset of size $C_g$  are identified as ``active'' , defining the current effective volume $V_g = V_o/C_g$.  During
this iteration, samples are drawn uniformly from this volume (e.g., uniformly from $V_o$ in iteration 0).
At any iteration $g$, the volume also has a specific number  $N_{\rm live}$ of live points $x_{g,\alpha}$, subdivided
among the $C_g$
cells with specific counts $n_{G}$
per bin $G\in [0,\ldots g]$.  The number of live points will usually increase steadily: points $x_\alpha$ are accumulated.

To start each iteration, the cell are resampled uniformly, preserving the number of live points within each cell,
producing a new set of live points $x_{g+1,\alpha}$.   The
likelihoods $L(x_\alpha)$ are evaluated, with points below the current likelihood threshold $L_{thr}$ immediately rejected.  Because our sampling prior is uniform, the empirical cumulative distribution of
$L(x_\alpha)$ estimates the true likelihood
cumulative.  On the one hand, with sufficiently many likelihood evaluations, we can estimate the fractional probability
associated with that live point (i.e., regions near that point), and reject points below any specific target likelihood.
Conversely, given a target probability threshold for coverage, we can estimate the threshold likelihood $L_{thr}$
needed to achieve that coverage level.   
During early iterations, when the likelihood distribution is not well-resolved, the code will simply include a fixed
large number $N_{min,*}$ of the highest likelihood points, and adjust $L_{thr}$ accordingly, until the estimated
``discarded probability'' is equal to $1-P_{thr}$.

\begin{figure*}
  \includegraphics[width=\textwidth]{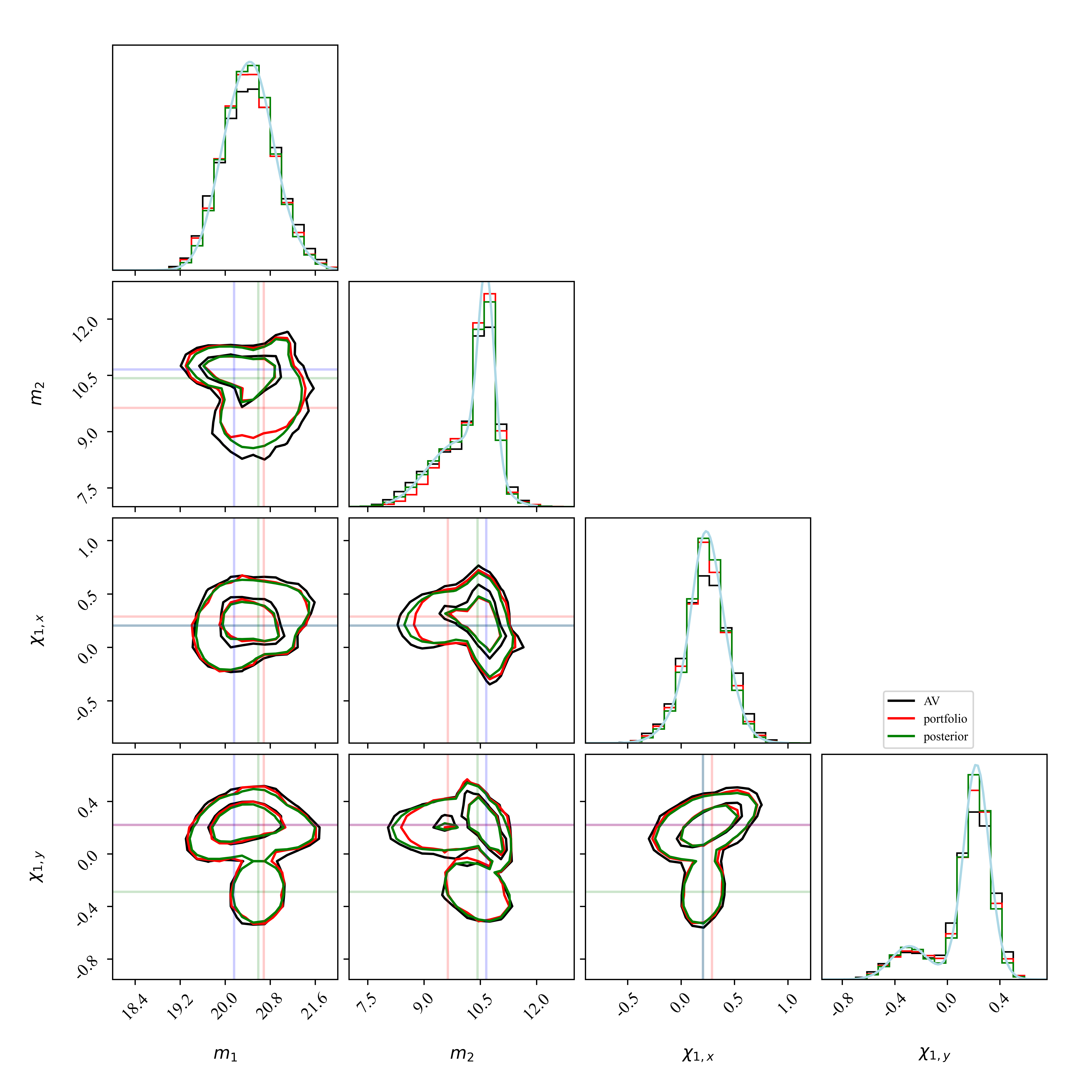}
  \caption{\label{fig:int:AV_toy}Demonstration of the adaptive volume (AV) and portfolio integrator in four dimensions, on three superposed multivariate
    gaussians with random relative weights, means, and  covariances.  The solid blue lines provide
    the exact one-dimensional marginals; the green lines and contours show the one- and two-dimensional marginal
    distributions estimated using draws from the posterior; and the black contours show the inferred posterior
    distribution produced by the AV integrator  using fair draws.   The red contours show corresponding results from a
    specific (and arbitrary) configuration of the portfolio integrator, described in the text.  To guide the eye, the three pairs of colored horizontal and vertical
    lines indicate the location of the three individual gaussian distributions.   This demonstration was
    performed with the production CIP code, calling an external likelihood function as described in Sec. \ref{sec:sub:external_lnL}, and performed in $\mc,\delta=(m_1-m_2)/(m_1+m_2)$ and
    cartesian spin coordinates.
    }
  \end{figure*}

Once identified, the likelihood threshold implies the number of live points above threshold, $N_{min}$; a Monte Carlo
estimate for the fractional volume associated with the live volume, 
$V_{g+1}=V_g N_{min}/N_{live}$; and the Monte Carlo uncertainty in this live volume $\Delta V_{g+1} = V_g \sqrt{N_{\rm
    min}}/N_{live}=V_{g+1}/\sqrt{N_{min}}$.
Refinement involves selecting a refinement scale: a revised number of bins in each direction, such that $1/n_{bins}^d$
is equal to the uncertainty in the \emph{total} sampled volume:
\begin{align}
\label{eq:nbins_V}
n_{\rm bins} = (1/\Delta V_{g+1})^{1/d} 
\end{align}
Critically, this expression regulates refinement and cell size:  each cell is of order the typical sampling-limited uncertainty
in the total  volume $V(L_{thr})$, preventing over-refinement.
Hypercubes in the refined grid are identified.
Finally, the integral and $n_{ESS}$ can be computed  using the weights:
\begin{align}
w_\alpha &\equiv L(x_\alpha)/\text{max}_\alpha L(x_\alpha) \\
I &\equiv \int L dx \simeq \frac{V_g }{N_{live}} L_{max} \sum_\alpha w_\alpha \\
n_{ESS} &= \frac{(\sum_\alpha w_\alpha)^2}{\sum w_\alpha^2}
\end{align}
As a first demonstration to show our port of the original VARAHA code preserves its correctness, Figure \ref{fig:int:AV_toy} shows a corner plot comparing the one- and two-dimensional marginal posterior distributions
for a superposition of two (randomly selected) multivariate Gaussians, on the one hand computed exactly and on the other
obtained using the AV integrator over a large region enclosing this narrow peak.

Our GPU-optimized implementation of the AV integrator is extremely efficient, with high sampling efficiency and reduced
cost compared to our previous implementations.  For the two-dimensional Rosenbrock toy problem, Figure
\ref{fig:n_eff_demo} shows the AV integrator achieves dramatically improved sampling efficiency.  In tests on real
events and practical ananlysis configurations, the typical cost per marginalized likelihood evalution using a full suite of
higher modes ($\ell_{\rm max}=4$) is roughly 2 seconds (which includes waveform generation), split equally between integration time (on the GPU) and
precomputation costs (on the CPU) to compute $U,V$; see Appendix \ref{ap:timing} for a concrete timing breakdown using
examples drawn from this work.  Even for the exceptional fiducial BBH shown in Figure
\ref{fig:fiducial_basic} (including the extra overhead of distance marginalization), we only needed 25 seconds per
marginal likelihood evaluation.   Compared to the benchmarks described in Appendix B of \cite{gwastro-RIFT-Update}, and
keeping in mind the new benchmarks reported above includes another factor of two speed improvement described below in
Section \ref{sec:technical_internal}, the AV integrator delivers roughly an order of magnitude improvement for typical problems.

\subsection{Portfolio integrator}
In practice the AV integrator provides more than enough flexibility for the configurations encountered in typical GW PE,
including inference of LISA sources.  Nonetheless, to protect RIFT's utility for the long term,
given  multiple strategies available for sampling, each with different strengths and weaknesses, we implemented a new
integrator to simultaneously employ  all of them.  
%
Leveraging python's plugin framework, this portfolio integrator also allows the user to employ any externally-developed
integrator software meeting RIFT's integrator interface specification, either individually or as a member of a larger portfolio.   

The new
portfolio integrator has a sampling prior which is a
(potentially dynamic) weighted average of the sampling priors $p_{s,\alpha}(x)$ derived from each integrator's sampling prior:  $p_s(x) =
[\sum_\alpha W_\alpha p_{s,\alpha} (x)]/\sum_\alpha W_\alpha$.  
The portfolio of sampling priors can be produced by the same sampler with different adaptation settings, or different
samplers entirely.  
Usually though not always, each sampler is trained using information provided by all other samplers' outputs.
The portfolio model allows us to provide a high-level framework to implement a wide variety of integration strategies
at the user's request.

As implemented for use within ILE and CIP, the current operation of the portfolio model currently provides a very narrow
range of user-accessible choices.   While end users can request a specific portfolio composition by (list of) $N_s$
names,
a list of options unique to each sampler realization, and  a list of $N_s-1$ integers specifying the first iteration for which a given sampler is activated within the
portfolio, 
the composition and settings of the portfolio is subsequently frozen within CIP or ILE.    As desired,
portfolio members can be configured contribute silently, only contributing  training data for other samplers without
directly influencing the integral.  For members contributing to the integral, the portfolio integrator will adapt
$W_\alpha$ based on the (instantaneous, per-sampler) effective sample size $n_{\rm ESS}$ seen in the last batch, more
highly weighting integrators which produce the highest $n_{\rm ESS}$ on the last batch of samplers.   These weights will
evolve but can never drop below $0.05/N_s$ where $N_s$ is the number of members of the sampler.  Normally, each
sampler's weights $W_\alpha$ and internal sampling prior $p_{s,\alpha}$ will be trained based on all the samples
in the training history (e.g., the last batch of samples), as each sampler's training routine is provided both those
samples $x_k $and the overall weights $w_k = {\cal L}(x_k) p(x_k)/p_s(x_k)$ associated with each sampler.   However, if any sampler
in the portfolio temporarily has weight  below $5\%$, the sampler is temporarily frozen: its training routine isn't called.
The many ad hoc numerical choices provided above have been chosen motivated by performance  a small suite of test
problems.

Relative to the already-powerful AV integrator, we anticipate that the portfolio integrator will be most useful in extreme situations, such as multiple comparable-magnitude but
narrow and non-overlapping local extrema (well
captured by a GMM) superimposed upon a broad background with complex structure (well captured by AV). 
To illustrate the portfolio technique with a more visually accessible example, the red curves in Figure \ref{fig:int:AV_toy}
show inference of a three-component gaussian mixture mdoel, using an arbitrarily-selected portfolio of two
conventional (AC) integrators with two different choices for their (fixed) resolution (10 and 30 bins, respectively).
We likewise make the ad-hoc choice that the high-resolution integrator will only start adapting its sampling after 10
iterations. After performing ten times as many likelihood evaluations as the AV integrator alone (also shown),
this portfolio converges to a comparable estimate of the posterior (e.g., both results have $n_{\rm ESS}\simeq 13,000$).

The portfolio integrator provides future RIFT users a path forward for future challenges in Monte Carlo inference.  Its
flexible structure potentially enables  simple implementations of widely-used techniques.  For example, with adaptively
chosen breakpoints, the portfolio integrator could conceivably perform simple nested sampling, particularly by using
Gaussian mixture models.   We emphasize however that while the portfolio integrator will perform better than its
components in general, it is not yet well-developed enough to be a panacea.  For example, naively  including a single-component GMM integrator
instead of one of the AC integrators in the analysis of  Figure \ref{fig:int:AV_toy}  produces results which miss one of the extrema.


\subsection{Unreliable oracle samplers}
The portfolio integrator's design allows us to incorporate what we call ``unreliable oracle'' (UO)  samplers: algorithms which
produce revised samples (based on either the likelihood model, previous training data, or both) for which we do not
precisely know the sampling distribution $p_s$.    These UO samplers include  random draws with replacement from
a fixed reference set; hill-climing maximization
algorithms; ``puffball''-style algorithms which jitter some of the training data; and a wide variety of  MCMC
methods, evolved briefly.    Lacking precise knowledge of the sampling distribution $p_s$ for UO, we always use UO
samplers as supplementary algorithms within portfolios: they provide insight to seed and stabilize our well-behaved
adaptive samplers.

Below, we briefly describe a few UO implementations distributed with RIFT.

\noindent \emph{Resampling oracle}: The resampling oracle returns a random realization of the input
samples $\{c_k\}$, supplemented as needed by randomly generating information for any dimensions not provided with the
input.   The resampling oracle provides a mechanism for RIFT to use previously-generated extrinsic or intrinsic
parameter inferences (e.g., skymaps or lower-dimensional PE) to initialize and accelerate sampling.   [Experienced RIFT
  users aready know that RIFT can be run  using an initial grid provided by another analysis, the ``bootstrap'' grid method.]

\noindent \emph{Puffball oracle}: The puffball oracle returns random draws from the mean and covariance of
(instantaneous) training data, truncated to the sample range.

\noindent \emph{Climbing oracle}: Our hill-climbing oracle uses randomly-selected high-likelihood points from the
training data to initialize a maximization algorithm.  Due to the high computational cost of maximization, this oracle
is designed to only be used to produce a small number of pure samples, optionally supplemented by nearby points.  This
oracle should be particularly helpful early on, to help initialize our robust sampling algorithms. 

\subsection{Normalizing flow integrator}
Though extremely efficient in low dimensional problems, the AV integrator has increasingly reduced sampling efficiency in higher dimensions (e.g., 5-8), due to its
fundamentally cartesian adaptive refinement method; see the discussion in Section \ref{sec:sub:comments}.  To
future-protect RIFT against the need for high sampling efficiency and low-latency analysis, we deploy an initial version
of a RIFT Monte Carlo integrator using normalizing flows (henceforth abbreviated as the NF integrator).

Normalizing flows are (a sequence of) one-to-one coordinate transformations whose inverse and determinants are known,
typically with (many) tunable parameters allowing the flow to mimic an arbitrary distribution \cite{pmlr-v37-rezende15,JMLR:v22:19-1028}.  Not long after their
inception, several groups recognized the potential of normalizing flows to facilitate high-performance multidimensional
Monte Carlo integration; see, e.g,. \cite{2020arXiv200105486G} and references therein.  
In parallel, several other groups in gravitational physics and elsewhere recognized the potential for normalizing flows
to improve other strategies for Bayesian inference
(e.g., using  normalizing flows for improved MCMC or nested sampling, 
\cite{2021PhRvD.103j3006W,gwastro-mergers-nessaiv2-2022,2022arXiv221106397W,2023ApJ...958..129W}  or directly for inference
\cite{2021PhRvL.127x1103D,gw-astro-mergers-dingo-part2,2022NatPh..18..112G}).  

We use the \texttt{nflows} package \cite{nflows} to generate and train normalizing flows based on weighted Monte Carlo
samples.
While in principle providing
higher sampling efficiency, at present due to costly re-optimization of
the flow, at present this code operates more slowly than our other integrators.    We anticipate that incremental code
improvements to access the backend libraries and choose suitable flow operating points will significantly improve the NF
integrator's wallclock performance over time.
Most important, however, the NF and portfolio integrators provide end-users with alternative tools to assess novel
results and investigate inevitable challenges that arise during discovery. 


\subsection{Integrator Validation and recommended operating point}
While Figure \ref{fig:int:AV_toy} is a visually convenient anecdote, as in previous work \cite{gwastro-RIFT-Update}, all integrators are validated against a small suite of standard
modest-dimension test problems, some of which are distributed with RIFT's source code as part of our
continuous-integration testing.
%
Later, we will describe end-to-end tests of our inference framework using the AV integrator in both ILE and CIP, both
for reference sources and in PP tests. 

To summarize, the AV integrator is sufficiently mature, lightweight, and useful for most low-dimensional inference
challenges. Particularly due to its long history of use in short-author work, demonstrating stability, we recommend it
as the sole integrator for ILE and CIP.  The portfolio integrator and NF integrator provide a promising options for
future investigation, but either introduce multiple  choices which each individually require careful validation (portfolio)  or are not yet optimized for
production use (NF).



\section{Technical advances: Other}
\label{sec:technical_internal}

\subsection{Reduce duplication of inner products}
RIFT's likelihood calculation relies on a factorized expression of the gravitational wave strain, using inner products
of spherical harmonic modes with the observational data (and other harmonic modes) to build the likelihood for generic
extrinsic parameters.    Previous studies have discussed how to use hardware acceleration to optimize two key  loops in RIFT's marginalization over
extrinsic parameters
\cite{gwastro-PENR-RIFT-GPU}: on the one hand, the marginalization of the likelihood over time, and on the other
marginalization over extrinsic parameters.  In this work, we revisit other elements of the calculation, particularly
associated with preprocessing, to further lower per-waveform evaluation cost.

When constructing the likelihood for a fixed $\bm \lambda$, RIFT first evaluates $U$ for one of the detectors involved,
to identify the subset of modes which contribute significantly to the overall likelihood.  Subsequent calculations of
$Q,U,V$ for all instruments then use a reduced mode set, significantly reducing overall cost for configurations where
only a few modes dominate.

For binaries with many higher-order modes ($\sum_\ell (2\ell+1) \simeq L_{max}^2$), the two inner product scalar-valued arrays $U$ and
$V$ are much more costly than the timeseries $Q$ to produce, due primarily to the  $O(L_{\rm max}^4)$ inner products
that must be evaluated.  Fortunately, 
for each interferometer, the matrices $U_{k,lm,l'm'} \equiv \qmstateproduct{h_{lm}}{h_{l'm'}}$ satisfy symmetry conditions, derived from the Hermetian inner
product:
\begin{align}
U_{k,lm,l'm'}^* = U_{k,l'm',lm} \\
V_{k,lm,l'm'} = V_{k,l'm',lm}
\end{align}
These symmetry relations allow a factor of 2 reduction in the number of inner product calculations needed.

In principle, we can also further reduce the need for inner products by constructing $Q_{lm}(\tau)$ over the necessary
narrow time window using time-domain operations.  For example, we can construct the time-domain signal corresponding to
$S_h(f)^{-1}d(f)$ once and for all, then convolve it for all times in our narrow analysis window with $h_{lm}(t)$.   In
practice, however, because the fast fourier transform scales as $N\log N$ with data length $N$, we perform comparably many
operations using fourier methods as a manual loop over coincidence times: only for the largest $N$ 
would this optimization be fruitful.

\subsection{Disable automatic mode filtering}
RIFT contains a subtle internal threshold $\epsilon$ (previously defaulting to $\epsilon=10^{-4}$, now defaulting to 0), which  on
a per-likelihood basis  filters the mode list $(\ell,m)$ based on whether any $|U_{lm,l'm'}/U_{22,22}|>\epsilon$ for some
$l'm'$.    For almost all low-amplitude signals with log-likelihood far below $1/\epsilon$, this restriction has minimal impact.  
If left unchanged, however, this threshold could bias inference of very loud or otherwise exceptionally-oriented sources. 
This filtering provided essential performance improvements in several early applications of RIFT and its precursors,
notably including direct comparisons to numerical relativity.   Becuase numerical relativity simulations report many
modes which have marginal practical data analysis impact, but the specific pertinent mode list will vary as the binary
total mass changes, this filtering choice allowed us to employ a large $L_{\rm max}$ and conservatively incorporate all
provided modes, but downselect in prractice to only the modes contributing to a specific analysis.

High-amplitude GW sources however are now much more frequent targets for real data analysis studies and particularly
investigations of future detector science.  For this reason, we have explicitly disabled this default, requiring the
user to manually invoke it if desired.

\subsection{More parameters in puffball}
As described in Section \ref{sec:sub:explore}, each iteration ends with RIFT providing at least two sets of samples.
The first set is our best estimate of the posterior distribution, as realized by many independent draws $\{\lambda_k\}$
from the distribution.  The second set  $\{\lambda'_k\}$ is derived from the first by randomly changing a subset of the
coordinates, based on the mean and covariance over those targeted coordinates, via the puffball routine.

Previously and as discussed above, RIFT would only jitter a specific subset of coordinates,
associated with the parameters dominating the leading-order orbital and gravitational wave phase: $(\mc,
\eta, \chi_{\rm eff})$.    Since these coordinates omit most spin degrees of freedom, in particular transverse spin, the
previous RIFT implementation could have adapting to explore the posterior distributions of binaries with large,
well-measured transverse spins.   To illustrate the importance of this parameter,  Figure \ref{fig:weak:spin_lobes} shows
analyses of a challenging synthetic event.
option. 
In the latest version of RIFT, at the end-user's request RIFT's puffball routine will  draw correlated random offsets for
$\mc,\eta,\chi_{i,z}$ as well as the transverse spin (the cynlindrical coordinates
$\chi_u,\phi$).
Unless otherwise noted, all RIFT analyses presented in this work use this transverse puffball technique. 


RIFT also previously  did not add random uncertainties to tidal parameters $\Lambda_i$.  When performing inference on
systems containing matter, we now always include tidal parameters to the list of parameters used by the puffball
routine.  Additionally, since $\Lambda_i$ must be non-negative, we have modified the puffball routine to perform its
covariance estimate and random offsets in $\ln \Lambda_i$, so that all jittered points have $\Lambda_i>0$. 

Sometimes, random offsets will produce unphyiscal outcomes in the puffball coordinate system, most notably scenarios
that lead to nominal $\eta>1/4$ or $\eta<0$.  These unphysical configurations were previously immediately rejected. To reduce boundary effects and increase efficiency, we now treat $\eta=0$
and $\eta=1/4$ as reflective boundaries inside the puffball calculation: offsets requested just outside this interval are
reflected (once) with the intent that the reflected point will lie within it.

\section{Results and examples}
\label{sec:results}

\subsection{Fiducial synthetic events}
Figure \ref{fig:weak:spin_lobes} shows inferences of two synthetic signals generated and interpreted with IMRPhenomXPHM using our preferred operating point: both ILE
and CIP use AV, while PUFF employs both aligned and transverse spin components.  To be concrete, both panels show
analyses of 4 seconds of data sampled at 4096 Hz, using a fiducial 3-detector
network characterized by synthetic Gaussian noise with presumed known PSDs, with a
Euclidean distance prior.  As noted previously, these fiducial signals have exceptional binary parameters (and SNR),
such that some or all parameters are well constrained.

\subsection{Analyses of selected O3 events with eccentricity}
\label{sec;sub:ex:ecc}
\begin{figure}
  \includegraphics[width=\columnwidth]{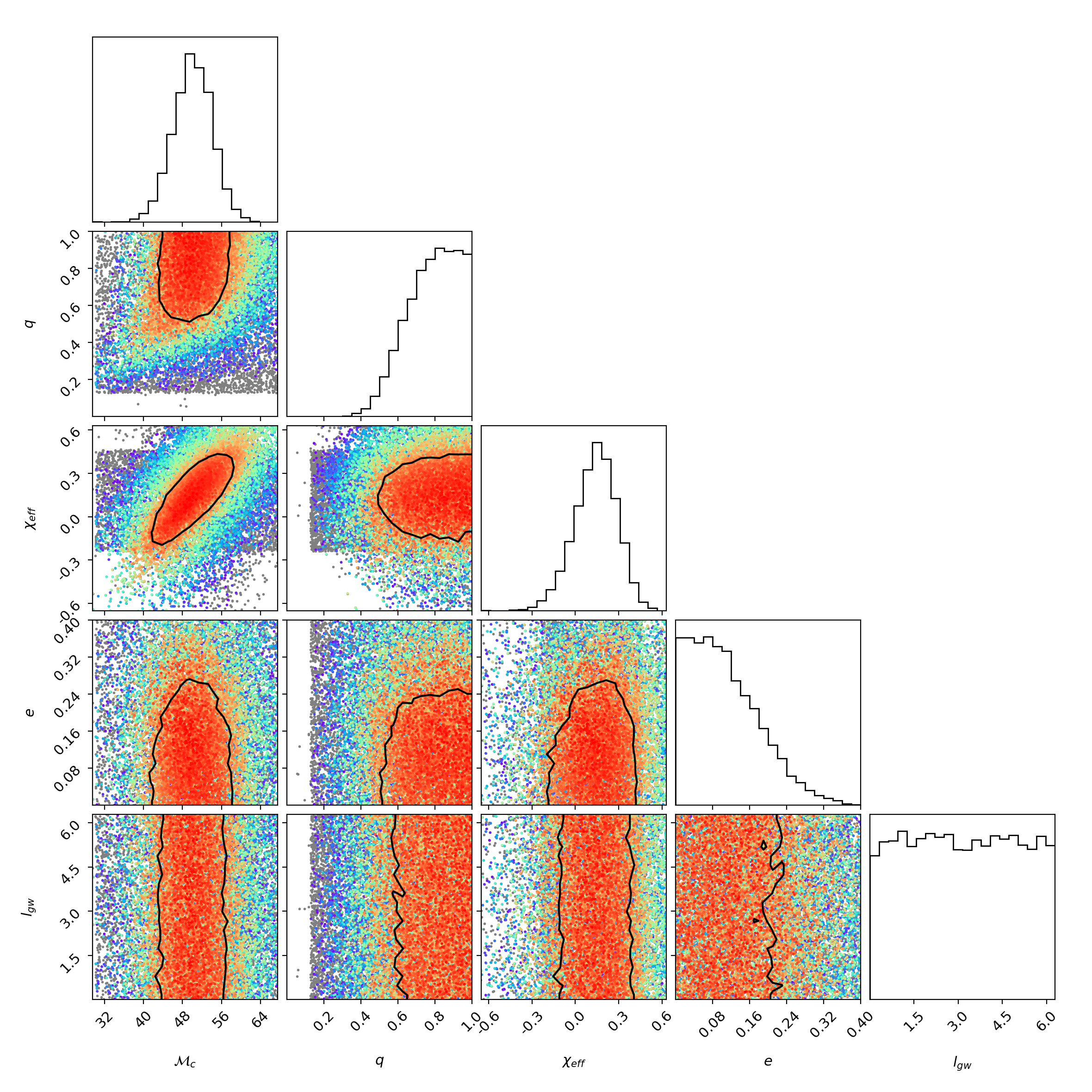}
  \caption{\label{fig:ex:asimov_v5EHM:gw200128}Example analysis of GW200128\_022011 with SEOBNRv5EHM
   }
\end{figure}

\begin{figure*}
  \includegraphics[width=0.9\linewidth]{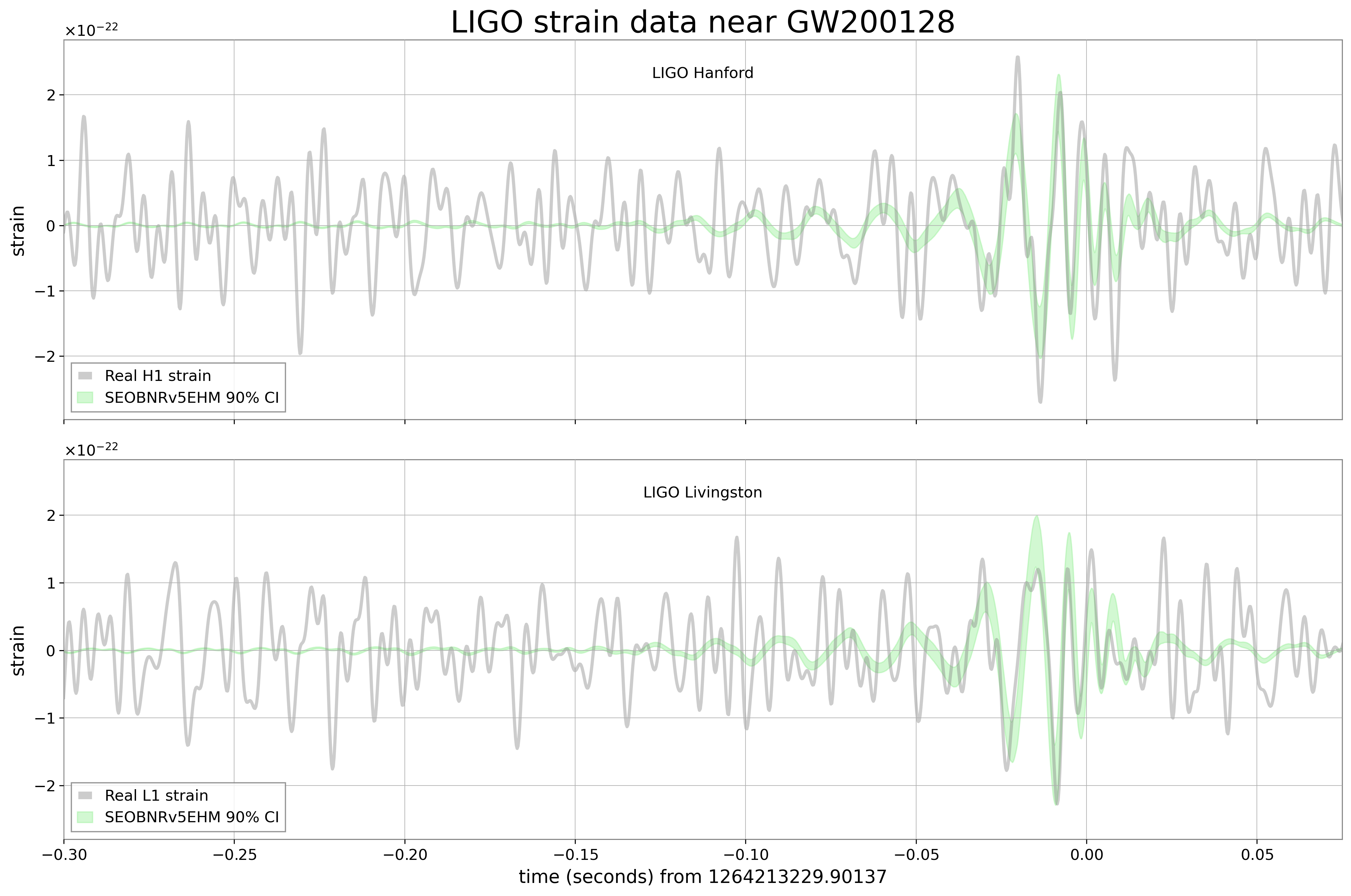}
  \caption{\label{fig:ex:ci_v5EHM:gw200128}Example 90\% confidence interval
    plot for an analysis of GW200128\_022011 with v5EHM.
   }
\end{figure*}

To demonstrate multiple new elements simultaneously -- asimov, eccentricity, the \texttt{gwsignal} waveform interface,
and our new integrators and operating points -- we reanalyze with an eccentric waveform model (SEOBNRv5EHM) several of the events previously studied with RIFT and asimov
\cite{gwastro-mergers-rift_asimov_O3-Fernando2024}.  For simplicity, we adopt a uniform prior on eccentricity between 0
and 0.5, with the upper bound motivated by the regime of validity of the waveform model.   Figure \ref{fig:ex:asimov_v5EHM:gw200128} provides
a corner plot showing our posterior marginal likelihood over intrinsic parameters, which for an eccentric binary include
the (model-dependent) eccentricity $e$ and mean anomaly $\ell$.   Figure \ref{fig:ex:ci_v5EHM:gw200128} shows the
corresponding confidence interval on the whitened signal, compared with the whitened data at this time.  Table
\ref{tab:asimov_v5EHM_keyparams} summarizes our inferences for the handful of events analyzed here.  All analyses are performed with our new default
operating point, which uses the AV integrator for all Monte Carlo integration, and assume perfect calibration.
For this proof-of-concept demonstration, we have explicitly omitted events with the strongest evidence for eccentricity
(i.e., GW200129\_065458); results on these events will be reported elsewhere.


\setlength{\tabcolsep}{6.6pt}
\renewcommand{\arraystretch}{1.6}
\begin{table*}
  \centering
  \begin{NiceTabular}{c c c c c c}
    \CodeBefore
    \rowcolors{2}{gray!25}{white}
    \Body
    Event & $\mathcal{M}_c (M_\odot)$ & $q$ & $\chi_{\rm{eff}}$ & $\mathcal{B}_{\text{E/QC}}$ & $e_{\text{GW}, 90\%}$ \\ 
    \hline
    GW191109\_010717 & $57.06_{-6.97}^{+6.36}$ & $0.74_{-0.20}^{+0.22}$ & $-0.35_{-0.21}^{+0.23}$ & $0.32$ & $0.23$ \\
    GW191222\_033537 & $50.46_{-7.84}^{+7.90}$ & $0.77_{-0.30}^{+0.20}$ & $-0.08_{-0.28}^{+0.25}$ & $0.31$ & $0.22$ \\
    GW191230\_180458 & $61.24_{-10.62}^{+9.32}$ & $0.79_{-0.28}^{+0.18}$ & $-0.07_{-0.37}^{+0.31}$ & $0.47$& $0.31$ \\
    GW200112\_155838 & $33.95_{-2.65}^{+2.89}$ & $0.79_{-0.24}^{+0.19}$ & $0.07_{-0.18}^{+0.17}$ &$0.20$ & $0.14$ \\
    GW200128\_022011 & $50.09_{-6.70}^{+6.37}$ & $0.80_{-0.26}^{+0.18}$ & $0.15_{-0.27}^{+0.22}$ & $0.38$& $0.25$ \\
    GW200208\_130117 & $ 38.57_{-5.55}^{+5.86}$ & $0.75_{-0.27}^{+0.22}$ & $-0.10_{-0.30}^{+0.27}$ & $0.29$& $0.19$ \\
    GW200209\_085452 & $41.72_{-8.00}^{+8.94}$ & $0.81_{-0.27}^{+0.17}$ & $-0.14_{-0.36}^{+0.31}$ & $0.47$& $0.31$ \\
    GW200216\_220804 & $60.36_{-15.26}^{+12.50}$ & $0.74_{-0.35}^{+0.23}$ & $0.09_{-0.60}^{+0.44}$ &$0.72$ & $0.42$ \\
    GW200219\_094415 & $42.13_{-6.94}^{+6.58}$ & $0.75_{-0.30}^{+0.22}$ & $-0.15_{-0.36}^{+0.29}$ & $0.41$& $0.28$ \\
    GW200220\_061928 & $128.32_{-30.63}^{+32.75}$ & $0.76_{-0.32}^{+0.21}$ & $0.28_{-0.65}^{+0.51}$ &$0.92$ & $0.36$ \\
    GW200220\_124850 & $46.73_{-8.40}^{+8.63}$ & $0.76_{-0.30}^{+0.21}$ & $-0.10_{-0.46}^{+0.36}$ & $0.58$& $0.39$  \\
    GW200224\_222234 & $40.86_{-3.77}^{+3.09}$ & $0.80_{-0.23}^{+0.18}$ & $0.13_{-0.16}^{+0.14}$ &$0.27$ & $0.19$ \\
    GW200302\_015811 & $29.75_{-4.25}^{+7.66}$ & $0.55_{-0.22}^{+0.37}$ & $0.03_{-0.31}^{+0.26}$ & $0.36$& $0.24$ \\
    GW200311\_115853 & $31.84_{-2.59}^{+2.76}$ & $0.83_{-0.25}^{+0.16}$ & $-0.07_{-0.19}^{+0.18}$ &$0.21$ & $0.14$ \\
    \hline
  \end{NiceTabular}
  \caption{Median and 90\% confidence intervals for key parameters from a set of runs performed using Asimov with
    v5EHM. For eccentricity, a one-sided upper bound is provided. }
  \label{tab:asimov_v5EHM_keyparams}
\end{table*}

\subsection{Automated cross-code comparisons of O3 events }
\begin{figure*}
  \includegraphics[width=0.48\textwidth]{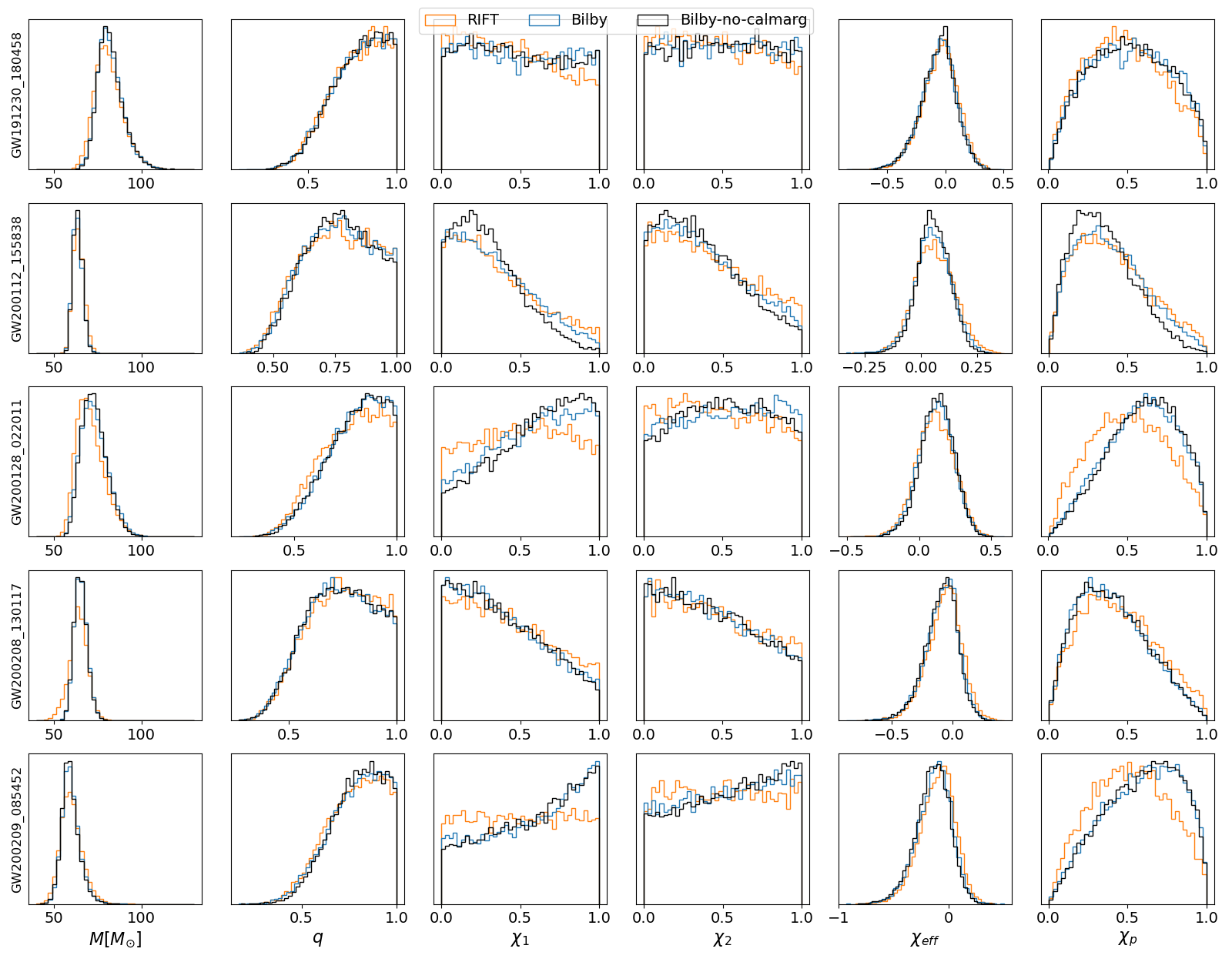}
  \includegraphics[width=0.48\textwidth]{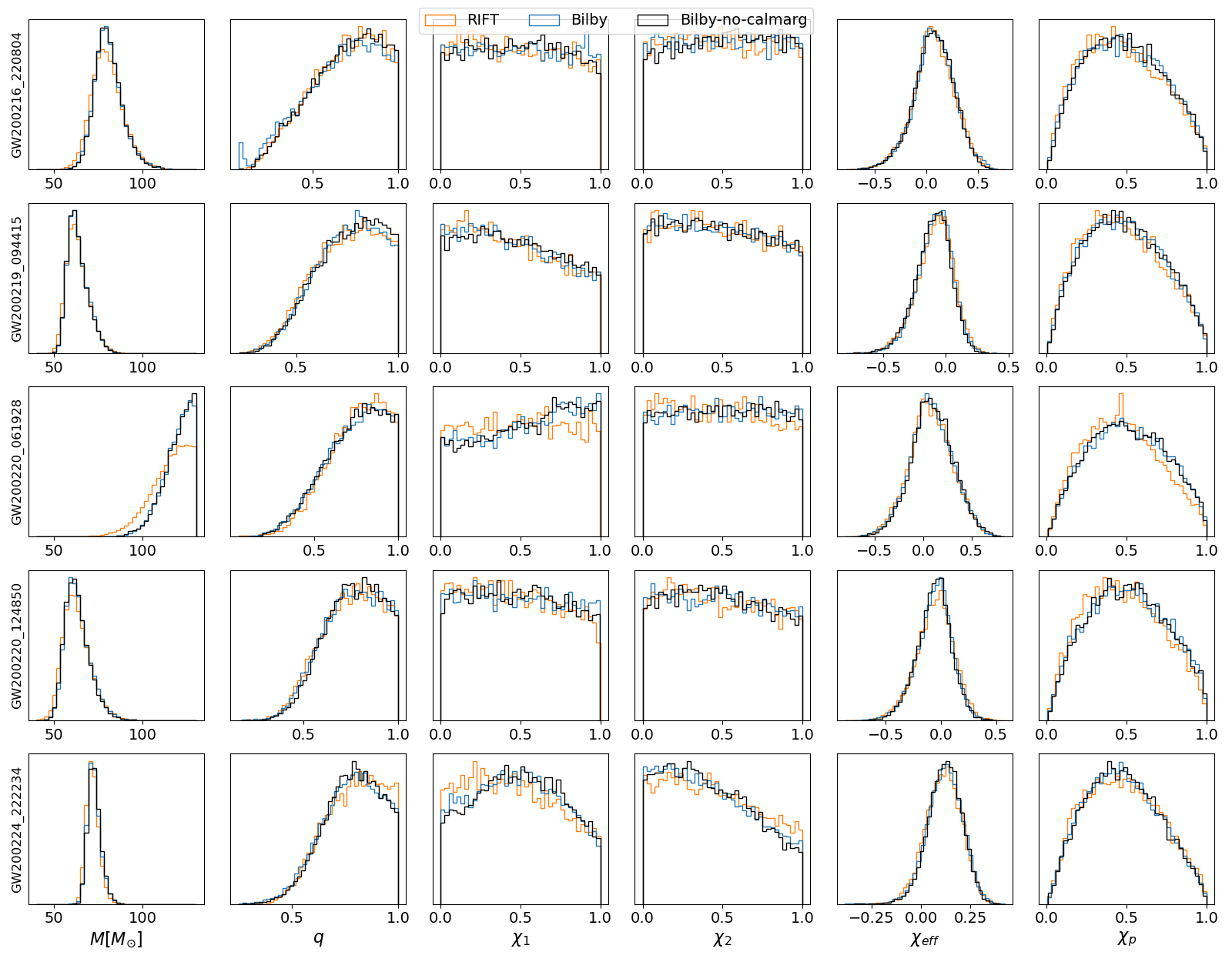}
  \caption{\label{fig:h2h:dee_events}Head to head comparison between RIFT and bilby on selected O3 events, using
    IMRPhenomXPHM.  In this example, the RIFT analyses (orange)  do not include the effect of calibration marginalization; bilby
    results are generated with (blue) and without (black) the effect of calibration marginalization.
    }
  \end{figure*}

Figure \ref{fig:h2h:dee_events}  shows a cross-code comparison between RIFT and bilby
analyses performed under the same settings used for \cite{gwastro-mergers-rift_asimov_O3-Fernando2024}, once again
neglecting calibration marginalization in the RIFT analyses. To ensure consistent asimov-compatible settings, we have
performed both the RIFT and bilby analyses ourselves, operating bilby using default configuration options inherited from
previous asimov analyses.  Both analyses use the corresponding default IMRPhenomXPHM implementation; see the discussion in Appendix
\ref{ap:hlm_conventions} for further discussion. All RIFT analyses use our preferred operating point: the AV integrator for both ILE and CIP,
with jitter applied to masses and both aligned and transverse spin components.  
For comparisons, we perform  two bilby analyses: one including and one omitting the effect of calibration
marginalization, using pre-existing settings for calibration marginalization envelope.  The two bilby runs are in
excellent agreement, suggesting calibration marginalization has little effect.  All RIFT and bilby analyses are performed
``out of the box'' using default settings, just as previous analyses with asimov
in \cite{gwastro-mergers-rift_asimov_O3-Fernando2024}, with no extra effort to assess convergence nor to further identify any residual
settings discrepancies.  While the two codes agree
reasonably well with these defaults,  modest differences for intrinsic parameters are apparent in some events, concentrated in configurations with notable skymap differences.
As this demonstration shows, head to head cross-code comparisons are now straightforward to generate over large sets of
events, potentially allowing faster and more pointed investigation into and diagnosis of minor differences between
inferred results derived from these two codes. 


\begin{figure}
  \includegraphics[width=\columnwidth]{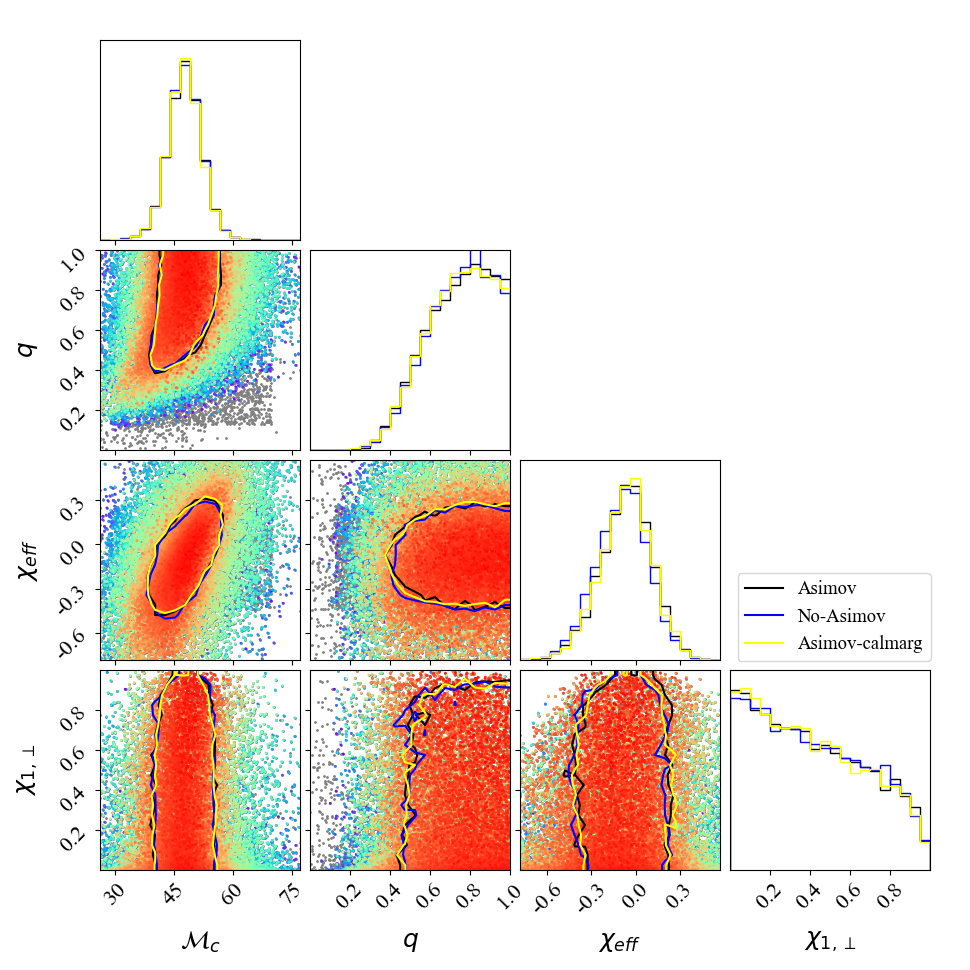}
  \caption{\label{fig:calmarg_demo}Analyses of GW200220 with multiple code configurations}
\end{figure}

To illustrate the stability of our results against using calibration marginalization and/or asimov, Figure
\ref{fig:calmarg_demo} shows the marginal posterior distributions derived using several different RIFT code
configurations, for a single event (GW200220). All agree.
Because calibration marginalization is only performed via postprocessing, all RIFT analyses natively
provide both a calibration-marginalized result and an analysis assuming perfect calibration, for comparison.

\subsection{Validation of AV via PP tests}
\begin{figure}
  \includegraphics[width=\columnwidth]{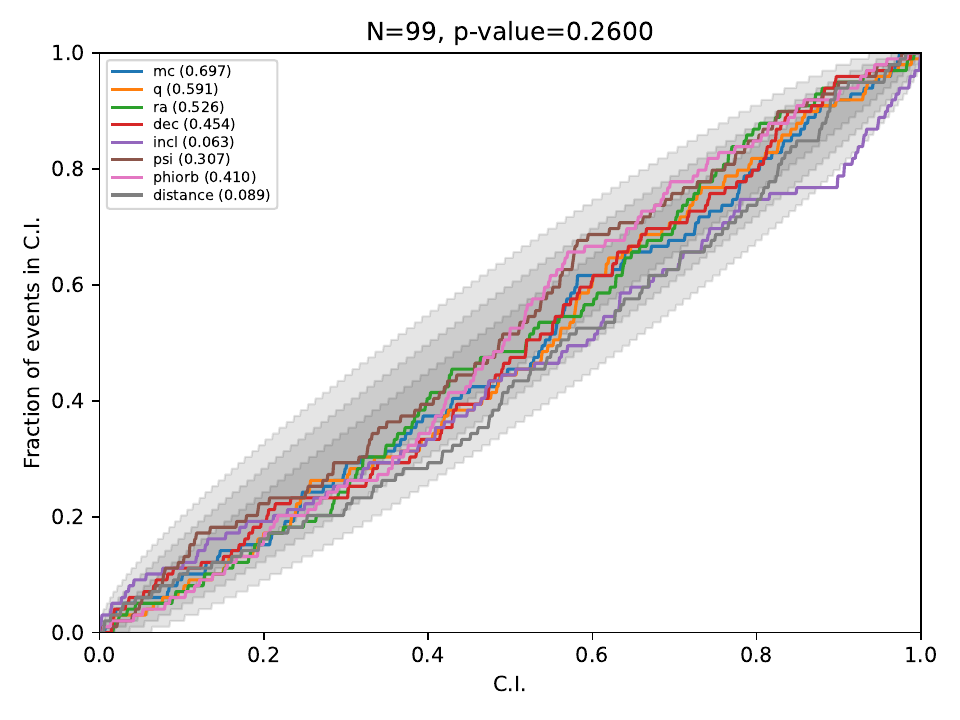}
  \caption{\label{fig:pp_0}\textbf{PP plot: zero spin and extrinsic}: Probability-probability plot to validate that the AV integrator
    operates correctly in ILE and CIP, using production settings throughout an end-to-end test.  Format follows the RIFT
    methods paper \cite{gwastro-PENR-RIFT}.  The synthetic sources and parameter
    inferences are constructed with IMRPhenomD \cite{Husa_2015iqa, Khan_2015jqa}
     in gaussian noise with presumed known
PSDs, for 3-detector networks, starting the signal at 20 Hz and using a 4096 Hz sampling rate, but only integrating the
signal up to 2 kHz.  Detector-frame masses are drawn uniformly in the region bounded by $\mc/M_\odot \in [10,20]$ and
$\eta\in[0.2,1/4]$. To produce sources with significant amplitude,  sources are drawn volumetrically between 800Mpc and
1500 Mpc.
}
  \end{figure}

\begin{figure}
  \includegraphics[width=\columnwidth]{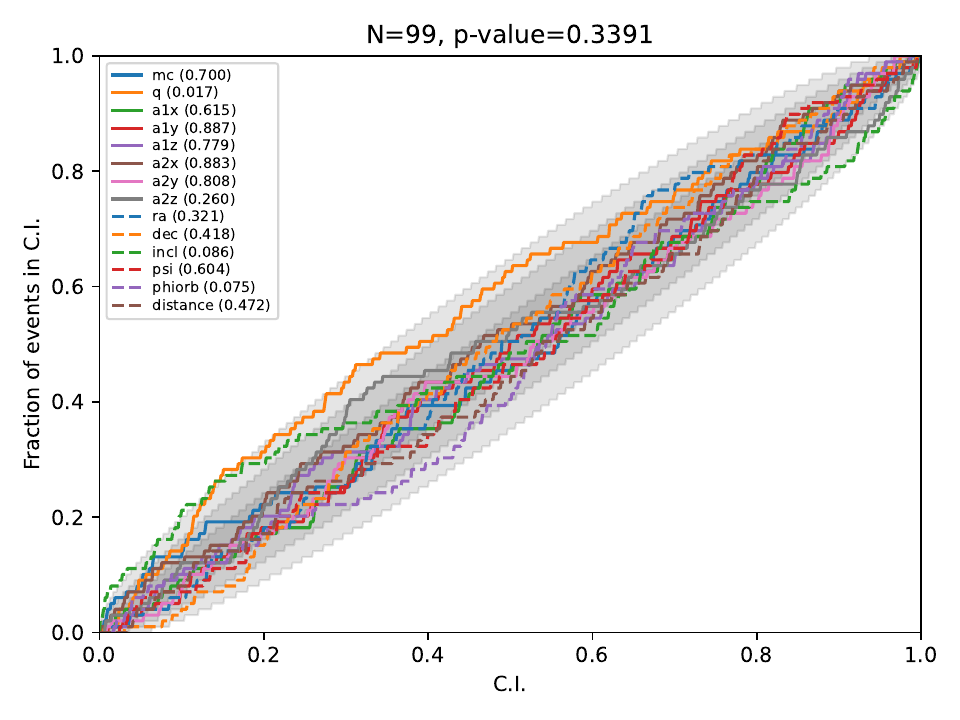}
  \caption{\label{fig:pp_Pv2}\textbf{PP plot: precessing spin extrinsic}: Probability-probability plot to validate that the AV integrator
    operates correctly in ILE and CIP, using production settings throughout an end-to-end test.  Format follows the RIFT
    methods paper \cite{gwastro-PENR-RIFT}.  The synthetic sources and parameter
    inferences are constructed with IMRPhenomPv2 \cite{Hannam_2013oca, Bohe_Pv2}
     in 16 seconds of gaussian noise with presumed known
PSDs, for 3-detector networks, starting the signal at 20 Hz and using a 4096 Hz sampling rate, but only integrating the
signal up to 2 kHz.  Detector-frame masses are drawn uniformly in the region bounded by $\mc/M_\odot \in [20,30]$ and
$\eta\in[0.1,1/4]$. To produce sources with significant amplitude,  sources are drawn volumetrically between 1Mpc and
6000 Mpc.
}
  \end{figure}

Following previous work \cite{gwastro-PENR-RIFT,gwastro-RIFT-Update}, we use end-to-end tests using randomly-drawn
sources to validate that our infrastructure and operating point work correctly with the new integrators used in ILE and
CIP.  For clarity of presentation and sharpness of credible intervals, we present these tests in incremental stages.
First,  Figure \ref{fig:pp_0} shows one such PP plot, using zero-spin sources randomly distributed over extrinsic
parameters to validate primarily that the AV works correctly for extrinsic parameter inference.
Next,  Figure \ref{fig:pp_Pv2} shows a PP plot using precessing sources, to highlight reliable recovery of binary black
hole parameters. 

\section{Conclusions}
\label{sec:conclude}
In this paper, we described several  improvements to RIFT's compact binary parameter inference method.  Using results of
previous studies to identify targeted limitations, we identified three areas for making RIFT more robust in practice:
better integrators and exploration; access to a wider family of waveforms; and bringing in-house followup calculations
like calibration marginalization and integration with asimov.   

To be concrete as regards the most noteworthy improvements,   we introduced an implementation of the AV Monte Carlo integrator; demonstrated its accuracy  by performing
controlled tests with known solutions; validated its
integration within the two key codes within RIFT which use it (ILE and CIP), within PP plots; and used it within our
proof-of-concept demonstrations.  Further, establishing a foundation for future growth, we also introduced three other
intgration-related tools: a normalizing flow integrator, the portfolio integrator, and unreliable oracles to guide
adaptive sampling.
We likewise introduced small but impactful technical improvements, such as allowing  RIFT to sometimes randomly jitter the precessing degrees of
freedom to better explore strongly-precessing signals.
Using multiple targeted examples, we demonstrated the latest version of RIFT resolves the challenges specificially
identified at the start of our study, while continuing to effectively employ cutting-edge waveforms to interpret
contemporary sources.

The improvements described here do not exhaust RIFT's capability for GW or generic inference.  For example, the
portfolio integrator alone provides numerous opportunities for dramatically improved efficiency, merely by tuning the
operating points or deploying improved integrators within it, or even adaptively tuning the portfolio dynamically based
on user-provided information.  As noted in the text, the portfolio mechanism can also be operated to mimic other
strategies such as nested sampling. The portfolio mechanism can even benefit from and directly employ other ongoing developments
in rapid GW parameter inference, such as normalizing flow methods (DINGO) and clever template-based placement \cite{2024arXiv241005190N}.
Conversely and more modestly, to increase sampling efficiency and reduce time to solution, successive iterations of
RIFT's integrators could benefit from modest guidance from previous versions applied to interpret the same GW data.
Finally, given already extremely efficient integration, RIFT would benefit from tighter optimization of its remaining components
to reduce redundancy, overhead, and excess cost.  Notably, RIFT could more efficiently perform self-avoiding grid
placement, to reduce the necessary grid size for robust solutions; better schedule its workers adaptively based on
measured need and performance; and employ faster tools to generate its
necessary inputs, including once-and-for-all file and PSD i/o, GPU-optimized waveforms and overlaps for $U,V,Q$, and
carefully tuned python imports to eliminate overhead.  As noted in the text, we anticipate RIFT's overall cost and time
to solution can still be reduced by several orders of magnitude.


\begin{acknowledgements}
  The authors would like to thank Michael Williams for helpful feedback on the
  manuscript. ROS and KW acknowledge support from NSF PHY 2012057.  ROS also
  acknowledges support from  NSF PHY 2309172 and 2206321. The authors are
  grateful for computational
  resources provided by the LIGO Laboratory and supported by National Science
  Foundation Grants PHY-0757058 and PHY-0823459. This material is based upon
  work supported by NSF's LIGO Laboratory which is a major facility fully funded
  by the National Science Foundation. This research has made use of data or
  software obtained from the Gravitational Wave Open Science Center (gwosc.org),
  a service of the LIGO Scientific Collaboration, the Virgo Collaboration, and
  KAGRA. This material is based upon work supported by NSF's LIGO Laboratory
  which is a major facility fully funded by the National Science Foundation, as
  well as the Science and Technology Facilities Council (STFC) of the United
  Kingdom, the Max-Planck-Society (MPS), and the State of Niedersachsen/Germany
  for support of the construction of Advanced LIGO and construction and
  operation of the GEO600 detector. Additional support for Advanced LIGO was
  provided by the Australian Research Council. Virgo is funded, through the
  European Gravitational Observatory (EGO), by the French Centre National de
  Recherche Scientifique (CNRS), the Italian Istituto Nazionale di Fisica
  Nucleare (INFN) and the Dutch Nikhef, with contributions by institutions from
  Belgium, Germany, Greece, Hungary, Ireland, Japan, Monaco, Poland, Portugal,
  Spain. KAGRA is supported by Ministry of Education, Culture, Sports, Science
  and Technology (MEXT), Japan Society for the Promotion of Science (JSPS) in
  Japan; National Research Foundation (NRF) and Ministry of Science and ICT
  (MSIT) in Korea; Academia Sinica (AS) and National Science and Technology
  Council (NSTC) in Taiwan.
  \end{acknowledgements}

\appendix

\section{Spherical harmonic frame conventions}
\label{ap:hlm_conventions}
The following appendix expands upon LIGO-T2300304, describing RIFT's frame convention.

RIFT uses the L  or radiation frame frame, in which the observer's viewing direction is
characterized by a vector direction $\textbf{n}$ \cite{gwastro-pe-systemframe} and this direction, along with $\hat{L}$,
defines an inertial frame's orientation.
The L frame defines the fiducial inertial frame such that $\hat{n}$ lies in the $x,z$ plane.  In terms of polar
coordinates  $\iota = \arccos(\hat{L} \cdot\hat{n})$, in this frame
\begin{align}
\hat{n} = \hat{z}_L \cos \iota +  \hat{x}_L \sin \iota
\end{align}
The spin components are specified in a frame aligned with $\hat{L}$, with the zero of the x axis chosen according to the
radiation frame.  The zero point of polarization is defined so $\psi=\psi_L=0$ is
aligned with the projection of $\hat{L}$ on the plane of the sky.
 Quasicircular binaries in the radiation frame \cite{gwastro-pe-systemframe}  are characterized
by their 15 L-frame intrinsic parameters $m_i,
\mathbf{\chi}_i$; their coalescence orbital phase $\phi_{\rm ref}$; the binary inclination $\iota$; the polarization $\psi_L$
corresponding to the instantaneous orientation of $\hat{L}$ on the plane of the sky; four spacetime coordinates
associated with the binary's distance and sky location.

 RIFT uses a complex gravitational
wave strain $h = h_+ - i h_\times$.  RIFT uses a spin-weighted
spherical harmonic decomposition in the L frame \cite{gwastro-PE-AlternativeArchitectures} with $\phi_*=0$:
\begin{align}
\label{eq:hlm}
h(t,\hat{n})  = e^{-2i\psi_L} \sum_{lm} h_{lm}(t|m_i,\textbf{S}_i) \Y{-2}_{lm}(\iota,\phi_{*}-\phi_{\rm ref})
\end{align}
In particular, RIFT assumes the $h_{lm}$ are independent of $\iota,\phi_{\rm ref},\psi_L$.

\begin{figure}
  \includegraphics[width=\columnwidth]{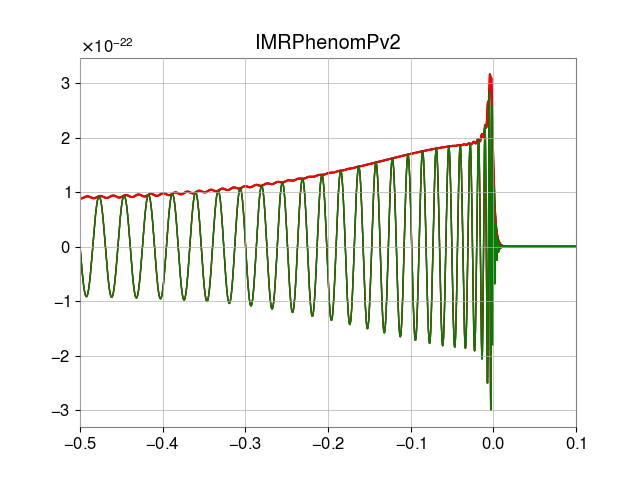}
  \caption{\label{fig:ap:check_imrpv2}
    Comparing the magnitude and real part of $h(\hat{n},t)$ generated for IMRPhenomPv2 (SpinTaylor spin evolution)
    for a fiducial precessing black hole binary, 
    generated using two code paths: the conventional implementation  (green and red) versus our method to generate
    $h_{2m}(t)$ from the same conventional implementation, five different times (black).
   The solid red and black envelopes show the amplitude $|h(t)|$.
    Due to the nearly-indistinguishable outcome of these two calculations, only the red (magnitude) and green (real
    part) are visible.
    }
  \end{figure}

RIFT obtains 
$h_{lm}(t)$ in this convention from a variety of codes, including \texttt{lalsimulation.ChooseTDModes} (or
ChooseFDModes, but see below), and \texttt{gwsignal} via \texttt{GenerateTDModes}.  These codes return modes with our
conventions (albeit with $\phi_*=\pi/2$).  Additionally RIFT provides a
workaround for a few models to extract $h_{lm}(t)$ from $h(t,\hat{n})$, described below.
RIFT creates $h_{lm}(t)$ for a very limited set of waveforms via two key workarounds:
\texttt{lalsim.SimInspiralTDModesFromPolarizations} (\href{https://git.ligo.org/lscsoft/lalsuite/-/blob/master/lalsimulation/lib/LALSimInspiral.c#L915}{source}) and \texttt{lalsimutils.hlmoft\_IMRPv2\_dict}.  Both methods extract
$h_k(t) \equiv h(t,\hat{n}(\theta_k,0))$ along a set of inclination angles $\theta_k$ and reconstruct $h_{lm}(t)$ using
assumptions about the mode content.  The first method only reconstructs $h_{2,\pm 2}(t)$ for nonprecessing binaries,
from $\theta_k=0,\pi$.
The second method reconstructs $h_{2m}(t)$ for precessing binaries with only quadrupole-order modes $L=2$ using
$\theta_k = k \pi/3$ for $k=0,1,2,3$  $\theta_4=\pi/2$, using the matrix ${\cal M}_{mk} \equiv \Y{-2}_{2m}(\theta_k,0)$.
By construction, these two workarounds produce $h_{lm}(t)$ which satisfies our conventions with $\phi_*=0$.
To illustrate the effectiveness of this workaround, Figure~\ref{fig:ap:check_imrpv2} compares the strain $h(t)$ derived using these two code paths, for a generic fiducial precessing binary.
Figure~\ref{fig:ap:check_imrpv2}  is a typical representative of the  excellent agreement between RIFT's modal approach and
the conventional implementation.

\begin{figure}
  \includegraphics[width=\columnwidth]{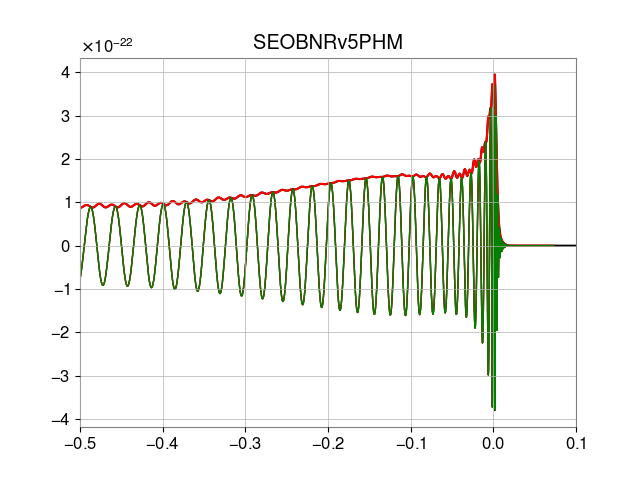}
  \caption{\label{fig:ap:check_SEOBNRv5PHM}
    Comparing the magnitude and real part of $h(\hat{n},t)$ generated for SEOBNRv5PHM
    for a fiducial precessing black hole binary, 
    generated using two code paths: the conventional implementation  (green and red) versus our method to generate
    $h_{\ell m}(t)$ from the modal representation and resum  (black).
   The solid red and black envelopes show the amplitude $|h(t)|$.
    Due to the nearly-indistinguishable outcome of these two calculations, only the red (magnitude) and green (real
    part) are visible.
    }
\end{figure}

\begin{figure}
  \includegraphics[width=\columnwidth]{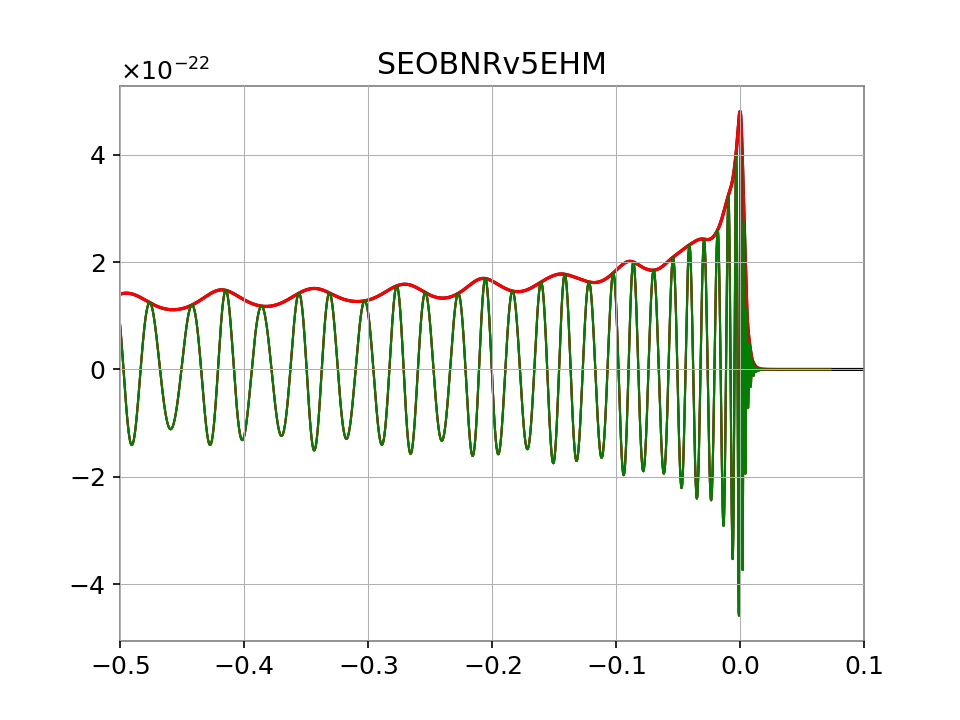}
  \caption{\label{fig:ap:check_SEOBNRv5EHM}
    Comparing the magnitude and real part of $h(\hat{n},t)$ generated for SEOBNRv5EHM
    for a fiducial precessing black hole binary, 
    generated using two code paths: the conventional implementation  (green and red) versus our method to generate
    $h_{\ell m}(t)$ from the modal representation and resum  (black).
   The solid red and black envelopes show the amplitude $|h(t)|$.
    Due to the nearly-indistinguishable outcome of these two calculations, only the red (magnitude) and green (real
    part) are visible.
    }
\end{figure}

The output of \texttt{gwsignal} conforms to RIFT's conventions with a phase  $\phi_*=\pi/2$.  The default settings of
RIFT's interface to \texttt{gwsignal} apply this phase shift to the modes natively returned by \texttt{gwsignal}. Figure
\ref{fig:ap:check_SEOBNRv5PHM} and Figure \ref{fig:ap:check_SEOBNRv5EHM} provide an example, generated for a random signal, validates agreement between our
phase-shifted mode convention and the  raw output provided by \texttt{gwsignal} for a specific interferometer's response.

\subsection{Working with J frame modes}
One class of waveforms provided by the \texttt{ChooseFDModes} routine return $h_{lm}$ in a different coordinate frame:
the $\hat{z}$ axis and polarization basis are associated with the $J$ frame.
When documenting the conventions for this approach, we closely follow Appendix C of \cite{gwastro-mergers-IMRPhenomXP} and the lalsuite documentation for ChooseFDModes.
According to that documentation, one can use these $\iota,\phi_{\rm ref}$-dependent $h_{lm}$ to
reconstruct the complex polarizations
\begin{align}
h_+ - i h_\times = e^{ +2i \zeta} \sum_{lm} h^J_{lm} \Y{-2}_{lm}(\theta_{JN},0) 
\end{align}
where $\zeta$ is a factor to correct for the orientation-dependent polarization (see below).

Physically, the transformation from J to $L$ frame has two obvious components to take $\hat{L}$ to the $\hat{z}$ axis:
$R(-\theta_{JN})R(-\phi_{JL})$.  The third angle $-\kappa$ rotates $\hat{N}$ to lie in the desired plane.  Their Euler
angle triplet is therefore $(\alpha,\beta,\gamma) = (-\kappa, -\theta_{JN}, -\phi_{JL})$, as seen in the appendix of \cite{gwastro-mergers-IMRPhenomXP}  Presumably because $\beta<0$ can cause
numerical problems with some implementations (square roots), they instead return $(\pi-\kappa, \beta,\pi-\phi_{JL})$,
which are related to the desired transformation by universal symmetry:
\begin{align}
e^{-i m\pi}d^\ell_{mm'}(-\beta) e^{-im'\pi} = (-1)^{m+m'}d^\ell_{m,m'}(-\beta) = d^\ell_{m,m'}(\beta)
\end{align}
In particular, they adopt the transformation in terms of their Euler angles $\alpha, \beta,\gamma$ of
\begin{align}
h^L_{lm} = \sum_{m'}  e^{i m' \alpha} e^{i m \gamma} d^\ell_{m',m}(\beta) h_{lm'}^J
\end{align}

The only non-obvious component of the calculation is the determination of $\kappa$ associated with rotating the observer
to the correct plane.
According to the XPHM docs, they adopt
a different fiducial observer in the $\hat{L}_0$ frame (i.e., for us, the L frame associated with our reference time)
\begin{align}
\hat{n}_{L,fid}& = \hat{z}_L \cos \iota + \sin \iota \times \nonumber \\ 
 & \left[\hat{x}_L \cos(\frac{\pi}{2} - \phi_{\rm ref}) + \hat{y}_L \sin(\frac{\pi}{2}-\phi_{\rm ref}) \right]
\end{align}
(In this expression, the choice of $\phi_*=\pi/2$ is manifest.) They also assume that in
the J frame, the oberver lies in the $x_J,z_J$ plane:
\begin{align}
\hat{n}_{fid} = \hat{z}_J \cos \theta_{JN} + \sin \theta_{JN} \hat{x}_J
\end{align}

\begin{figure}
  \includegraphics[width=\columnwidth]{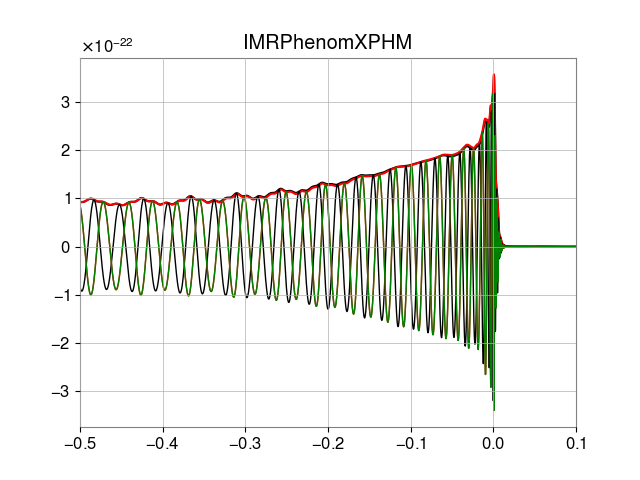}
    \includegraphics[width=\columnwidth]{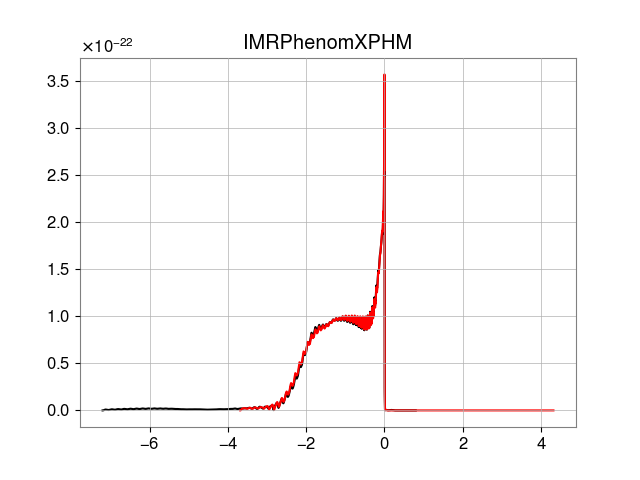}
  \caption{\emph{Top panel}: Comparing $h(\hat{n},t)$ generated for IMRPhenomXPHM (SpinTaylor spin evolution)
    for a fiducial precessing black hole binary, 
    generated using two code paths: the conventional implementation  (green and red) versus our modal reconstruction,
    including frame rotation to change from the $J$ to $L$ frame
    (black).   The solid red and black envelopes show the amplitude $|h(t)|$.  Except for an overall difference in
    polarization convention, the two implementations are in excellent agreement.
    \emph{Bottom panel}: Comparison of $|h(t)|$ for the same two code paths and parameters for a longer time range,
    illustrating small amplitude disagreements due to very small differences in data conditioning.
    }
  \end{figure}

To generate $h_{lm}(t)$ in the RIFT frame, with the desired polarization basis,  we generate the $h_{llm}^J$ using ChooseFDModes plus an inverse fourier
transform, using arguments with $\iota=\phi_{ref}=0$.  We use the Euler angles above, corrected\footnote{We empirically
  find this expression reproduces for example SEOBNRv4PHM precessional dynamics of the outgoing radiation} by $\alpha \rightarrow
\alpha'=\alpha+\pi$  and accounting for the polarization angle $\Delta \psi_{JL}$ associated with the projection of $J$
onto the plane of the sky defined by the radiation frame (assuming small positive $\iota$ to break any degeneracy).

\section{Exploiting redundancy for validation}
\label{ap:checks}
\noindent \emph{Calibration marginalization or not}: As noted in the text, RIFT performs calibration marginalization via
postprocessing.  First,  RIFT's raw output is re-assessed, assigning weights to each point; then, following LVK
conventions for output, RIFT performs a fair draw from the reweighted samples to produce unweighted fairly drawn
samples.  As usually calibration marginalization has almost no effect, the two sets of posterior samples should closely
agree, up to modest changes in the skymap.   This check allows us to test the impact of calibration uncertainty.

\noindent \emph{Before and after extrinsic resampling}: RIFT generates its final extrinsic samples from a final set of
intrinsic-only posterior samples: each such sample is passed to a fast run of ILE to produce one (or sometimes several) fair extrinsic
samples. Therefore, the intrinsic-only and extrinsic samples should agree.  This consistency test can diagnose whether
ILE sampling has been ineffective for some intrinsic configurations, which could for example  involve a particularly exceptional extrinsic configuration (e.g., an edge-on strongly
precessing binary, whose parameters are all tightly constrained).  Prior to adopting the AV integrator, some
exceptional combinations of intrinsic and extrinsic configurations were difficult to sample. 

\noindent \emph{Before and after convergence testing}: As normally operated, RIFT usually iterates to convergence with
relatively low-resolution sampling, merely sufficient for exploration and convergence testing.  Once converged, the full
posterior generation code is run with many CIP worker instances, to generate a full extrinsic posterior.  The
low-resolution and high-resolution sampling should agree.  This consistency test allows us to diagnose whether RIFT's settings are 
well-adapted to interpret a specific observation.

\noindent \emph{Overall and per-iteration marginal likelihoods: agreement and error}: The most common RIFT consistency test is checking whether posterior
credible intervals encose all regions of largest marginal likelihood, or otherwise providing a compelling reason why that should
not occur (e.g., strong priors).   In the best-converged analyses, the marginal likelihood colormap smoothly transitions in
scale over a region surrounding the posterior.  

\section{Timing analysis for marginal likelihood evaluation}
\label{ap:timing}

As in previous work \cite{Wysocki_2019}, we have pointed out that RIFT's marginal likelihoods are fast, low-cost
calculations.   In this work, we update information supplied in previously-reported benchmarks, to highlight RIFT's
performance and rate-limiting factors given on the one hand improved integrators and on the other the latest costly
time-domain waveform models.
Table \ref{tab:speedchar} provides a quantitative summary of RIFT's marginal likelihood evaluation cost, asessed by
profiling the ILE executable during the calculation of five different sets of intrinsic parameters drawn from the
respective posterior distributions.. As a reminder, the ILE
stage of RIFT computes a marginalized likelihood of the data over the intrinsic
source parameters. Each instance of ILE examines a
user-specified number of intrinsic points in parallel to best make use of
computing resources.   Aside from the cost incurred by starting up ILE itself -- loading the code and its libraries,
data, PSDs, and the list of intrinsic parameters themselves --  the cost for each intrinsic point includes a startup cost,
a setup cost, and an integration cost. We show this per-event total evaluation cost in the table
below as the ``overall'' cost of one instance of ILE. First, the dynamics and
outgoing radiation are computed for each set of instrinsic parameters in a
``precompute'' stage. This includes both waveform generation and the inner
products. Next, the likelihood can be efficiently calculated only as
a function of the extrinsic parameters in the ``likelihood'' stage. The
``integration'' stage refers to the Monte Carlo integration of the extrinsic
samples. To highlight the speed of the likelihood calculation, we also include a
column in the table describing how many points for which the likelihood was
calculated.  All likelihood evaluation is performed using time marginalization.

The table below shows profiling performance for some example runs from plots
shown previously in this paper. We choose a variety of setups, from very simple, to
very complicated, to show that RIFT performs well even in the most difficult
configurations. To see more information about RIFT's performance, see
\cite{Wysocki_2019}. The first two rows of the table profile an event taken from
the precessing PP plot show in Figure \ref{fig:pp_Pv2}. This event is a
synthetic source in 16 seconds of gaussian noise with presumed known PSDs for a
three detector network using a 4096Hz sampling rate. We profile this event with
both IMRPhenomPv2 and SEOBNRv5PHM. We find that SEOBNRv5PHM has a much more
expensive precompute stage, reflecting both the substantialy increased cost of generating a single waveform and the
larger number of modes used in our analysis.  The
middle two rows of the table show profiling for a difficult source, the fiducial
event shown in Figure \ref{fig:weak:spin_lobes}. This source is edge-on, with
transverse multi-modal spins, and has a high mass ratio. It was injected into 8
seconds of gaussian noise with presumed known PSDs for a three detector network,
also using a 4096Hz sampling rate. We profile this run with both IMRPhenomXPHM
and SEOBNRv5PHM.   Compared to the previous event using similar $\ell_{\rm max}=4$, more than four times as many likelihood evaluations are needed to
produce an acceptable final result; consequently, the overeall likelihood evaluation time is also several times larger.  Finally, the last event in the table is
GW170817, a BNS source. We analyzed 128 seconds of data with a 4096Hz sampling
rate, using a waveform that included tidal parameters, IMRPhenomPv2\_NRTidal.

In all cases, we see that the likelihood costs are very low.  In contrast to previous work, the precompute stage
is half or more of the overall time to solution in all cases.  While the
precompute stage that involves the waveform generation and inner products  could
be further optimized, we leave the necessary code updates to future work.

\begin{table*}
  \centering
  \begin{NiceTabular}{|c|c|c|c|c|c|c|c|c|}[vlines]
    \CodeBefore
    \Body
    \hline
    \makecell{Waveform} & 
    \makecell{srate\\(Hz)} & 
    \makecell{Duration\\(sec)} & 
    \makecell{$\ell_{\rm max}$} &
    \makecell{precompute\\(sec)} &
    \makecell{likelihood\\(sec)} & 
    \makecell{integrate\\(sec)} &
    \makecell{overall\\(sec)} & 
    \makecell{$N_{\cal L}$} \\
    \hline
    Pv2 & 4096 & 16 & 2 & 0.588 & 0.186 & 0.13 & 0.904 & 24,158 \\
    v5PHM & 4096 & 16 & 4 & 5.699 & 0.417 & 0.583 & 6.699 & 64,927 \\
    \hline
    XPHM & 4096 & 8 & 4 & 4.701 & 1.50 & 3.83 & 10.03 & 307,443 \\
    v5PHM & 4096 & 8 & 4 & 5.845 & 1.343 & 3.23 & 10.419 & 248,757 \\
    \hline
    Pv2\_NRTidal & 4096 & 128 & 2 & 7.351 & 0.157 & 0.113 & 7.621 & 28,137 \\
    \hline
  \end{NiceTabular}
    \caption{Profiling performance: evaluation costs for ILE using GPUs with the
      AV integrator. We include the waveform model, the sampling rate used for
      the analysis, the duration of the data, and the number of modes included
      in the profiling analysis. Next we split the overall cost of an ILE job
      into parts. These costs represent the time for a single set of intrinsic
      parameters. The precompute stage inlcudes waveform generation and inner
      products. The likelihood is computed for each set of extrinsic parameters,
      and we show the total number of extrinsic points in the last column,
      averaged over five sequential ILE computations (each with different
      intrinsic parameters) which is done for computational efficiency. The
      Monte Carlo integration cost is a derived quantity that occupies the
      remainder of the overall time. The sum of those three stages is the
      overall cost. We profile three sources - an ``easy'' BBH, a ``difficult''
      BBH, and a BNS. The first two rows are an example taken from the
      precessing PP plot shown in Figure \ref{fig:pp_Pv2}. The next two rows
      reflect the fiducial event shown Figure \ref{fig:weak:spin_lobes}. The
      last row is for BNS source GW170817. These sources are described in
      further detail in the text.}
  \label{tab:speedchar}
\end{table*}

\bibliography{LIGO-publications,gw-astronomy-mergers,gw-astronomy-mergers-approximations,textbooks,extra,gw-astronomy-detection,gw-astronomy-mergers-dynamical-agndisk,popsyn_gw-merger-rates,gw-astronomy-mergers-wd,gw-astronomy-mergers-ns-gw170817,gw-astronomy-mergers-ns-nuclearphysics,gw-astronomy-mergers-eccentric}
\end{document}